\documentclass[12pt, preprint]{aastex}

\usepackage{graphicx}
\usepackage{subfigure}
\usepackage[normalem]{ulem}
\usepackage{epstopdf}
\usepackage{color}
\usepackage{hyperref}
\usepackage{url}

\usepackage{natbib}
\usepackage[usenames,dvipsnames,svgnames,table]{xcolor}
\definecolor{darkgreen}{RGB}{0,100, 0}

\newcommand{\project}[1]{\textsl{#1}}
\newcommand{\Kepler}{\project{Kepler}}
\newcommand{\name}{CPM}
\newcommand{\set}[1]{\mathcal{#1}}

\definecolor{linkcolor}{rgb}{0,0,0.5}
\hypersetup{colorlinks=true,linkcolor=linkcolor,citecolor=linkcolor,
            filecolor=linkcolor,urlcolor=linkcolor}

\begin{document}

\title{
  A Causal, Data-Driven Approach to Modeling the Kepler Data
 \\ 
}
\author{%
  Dun~Wang\altaffilmark{\ref{CCPP}},
  David~W.~Hogg\altaffilmark{\ref{CCPP},\ref{CDS},\ref{MPIA},\ref{email}},
  Daniel~Foreman-Mackey\altaffilmark{\ref{UW},\ref{SF}},
  Bernhard~Sch\"olkopf\altaffilmark{\ref{MPIIS}}
  }
\newcounter{address}
\setcounter{address}{1}
\altaffiltext{\theaddress}{\stepcounter{address}\label{CCPP}%
  Center for Cosmology and Particle Physics, Department of Physics, New York University}
\altaffiltext{\theaddress}{\stepcounter{address}\label{CDS}%
  Center for Data Science, New York University}
\altaffiltext{\theaddress}{\stepcounter{address}\label{MPIA}%
  Max-Planck-Institut f\"ur Astronomie, Heidelberg, Germany}
\altaffiltext{\theaddress}{\stepcounter{address}\label{email}%
  To whom correspondence should be addressed; \texttt{<david.hogg@nyu.edu>}.}
\altaffiltext{\theaddress}{\stepcounter{address}\label{MPIIS}%
  Max-Planck-Institut f\"ur Intelligente Systeme, T\"ubingen}
\altaffiltext{\theaddress}{\stepcounter{address}\label{UW}%
 Astronomy Department, University of Washington, Seattle, WA 98195}
\altaffiltext{\theaddress}{\stepcounter{address}\label{SF}%
Sagan Fellow}

\begin{abstract}
Astronomical observations are affected by several kinds of noise, each with its own causal source; 
there is photon noise, stochastic source variability, and residuals coming from imperfect calibration of the detector or telescope. 
The precision of NASA \Kepler\ photometry for exoplanet science---
the most precise photometric measurements of stars ever made---%
appears to be limited by unknown or untracked variations in spacecraft pointing and temperature, and unmodeled stellar variability. Here we present the Causal Pixel Model (\name) for \Kepler\ data, a data-driven model intended to capture variability but preserve transit signals. 
The \name\ works at the pixel level so that it can capture very fine-grained information about the variation of the spacecraft.
The CPM models the systematic effects in the time series of a pixel using the pixels of many other stars and the assumption that any shared signal in these causally disconnected light curves is caused by instrumental effects. 
In addition, we use the target star's future and past (auto-regression). 
By appropriately separating, for each data point, the data into training and test sets, we ensure that information about any transit will be perfectly isolated from the model. 
The method has four tuning parameters---the number of predictor stars or pixels, the auto-regressive window size, and two L2-regularization amplitudes for model components, which we set by cross-validation. 
We determine values for tuning parameters that works well for most of the stars and apply the method to a corresponding set of target stars.
We find that \name\ can consistently produce low-noise light curves. In this paper, we demonstrate that pixel-level de-trending is possible while retaining transit signals and we think that methods like \name\ are generally applicable and might be useful for  K2, TESS, etc, where the data are not clean postage stamps like \Kepler.

\end{abstract}

\section{Introduction}

The photometric measurements of stars made by the \Kepler\ spacecraft are precise enough
  to permit discovery of exoplanet transits with depths smaller than $10^{-4}$.
This precision results from great spacecraft stability,
  supplemented by various methods for removing small residual spacecraft-induced and stellar-variability trends in the brightnesses,
  either filtering the data (with things like median filters)
  or fitting the data with flexible models (like polynomials or splines or Gaussian Processes; 
  \citealt{gaussian}).
When employed in the service of exoplanet search and characterization,
  these methods are usually agnostic about whether photometric variations originate in the spacecraft or in the star itself;
  that is, they obliterate intrinsic stellar variability along with spacecraft issues.

In general, there are many reasons for apparent photometric variability in a \Kepler\ source.
There is intrinsic stellar variability,
  which is of interest to some and a nuisance to others.
There is also variability of overlapping fainter stars;
  that is, confusion noise combined with variability of the confusing sources.
There are small changes in spacecraft pointing,
  which leads to slightly different illumination of the focal-plane pixels,
  and thus different sensitivity to variations in the device.
There are also \emph{intra-pixel} sensitivity variations that can contribute(\citealt{subpixel}).
There are small changes in spacecraft temperature,
  which lead to point-spread function (PSF) and differential pointing changes.
These also lead to changes in pixel and intra-pixel illumination.
Stellar proper motion, geometric parallaxes, and differential stellar aberration as the spacecraft orbits all do more of the same.
There is electronic cross-talk between detectors and charge-transfer inefficiency;
  these can effectively transfer variability from one source to another.
There are additional electronics effects like ``rolling bands'' that put additional features into light curves \citep{handbook}.
There are also changes to the detector sensitivity with temperature and time,
  possibly illimination-history effects,
  and possibly other sources of variability not yet considered.
The remarkable thing about \Kepler\ is that it is trying to measure stars at a level of precision
  much higher than ever previously attempted;
  new effects really \emph{must} appear at some point.
In \figurename~\ref{ccd}, we show the pixel-level variability in the \Kepler\ data
  near one bright star that shows minimal intrinsic variability;
  this figure highlights the spacecraft-induced effects.

We propose to mitigate these variations in \Kepler\ light curves by \emph{modeling} them.
In this context, a model is a parameterized function that can predict data, given parameter settings,
  and an objective function that can be used to set or sample those parameters.
In some cases, the objective function can be given a probabilistic justification or interpretation, e.g., if it is constructed using a likelihood and a prior pdf.
The model can be a physical model (of the spacecraft PSF, pointing, temperature, and so on)
  or it can be a flexible, effective model that has no direct interpretation in terms of physical spacecraft parameters.
If done well, we would expect a physical model to do a better job,
  because it embodies more prior information,
  but it requires research and intuition about dominant effects. 
If this research and intuition is wrong or incomplete, a physical model may actually perform worse.
The model we propose here is in the non-physical, effective category.
The \Kepler\ community is familiar with these kinds of models; to our knowledge, \emph{all} successful light curve ``de-trending'' methods are flexible, effective models.

One such method---one that is designed to describe or model spacecraft-induced problems
  but \emph{not} interfere with measurements of stellar variability---%
  is the \Kepler\ Presearch Data Conditioning (PDC, \citealt{pdc1}).
The PDC builds on the idea that systematic errors have a temporal structure that can be extracted from ancillary quantities. 

In the first PDC \citep{pdc1}, 
  removal of systematic errors was performed based on correlations with a set of ancillary engineering data. 
These data include the temperatures at the local detector electronics below the CCD array, 
  and polynomials describing the centroid motion of the targets from PA(Photometric Analysis).

In the most recent version of PDC (PDC-MAP, \citealt{pdc2,pdc3}), filtered light curves of other stars are used. Using a set of relatively quiet stars on each detector, the top 8 principal components are computed for every quarter of Kepler data. 
The maximum a posteriori linear combination of these basis light curves---computed using an empirical prior on the trend magnitudes---are used to model and remove the systematics in all the Kepler light curves. 
The choice to use such a small number of components and to apply priors was made in order to reduce over-fitting and retain physical signals after systematics removal.

PDC ``co-trends'' or calibrates the \Kepler\ light curves by removing those parts that are explained by the basis light curves generated from a principal components analysis (PCA) of filtered light curves.
That is, it exploits statistical dependences between different time series, regularizing the fit (and avoiding over-fitting) 
  by filtering and restricting the dimensionality (through PCA).

The method proposed here, The Causal Pixel Model(\name), shares the motivation of PDC.
The main differences are that
\begin{enumerate}
\item
\name\ works in the pixel domain, not the light curve domain, so it has access to more fine-grained information
\item
\name\ has far more freedom (far more parameters) than the PDC but it strictly avoids over-fitting the light curves on exoplanet-transit time-scales through strong regularization and a train-and-test framework. 
PDC mitigates overfitting by
reducing the freedom of the model and applying empirical priors to the
fit coefficients.
\item 
\name\ directly optimizes for prediction accuracy, while PDC uses PCA projections.
\item
The current version of \name\ is
tuned to remove systematics while minimizing the impact on exoplanet transit signals. 
It makes no attempt to retain lower frequency signals that can be produced by other astrophysical signals.
\item 
\name\ uses as inputs only instantaneous values of other stars, plus near future and past of the star itself in an autoregressive fashion, while PDC works with time series (light curves) of the other stars.
\end{enumerate}

The two methods (\name\ and PDC) are similar however, in that they both make the assumption that whatever spacecraft effects are imprinting variability on a stellar light curve must also be imprinting similar or related variability on other light curves.
By going to the pixel level (unlike the PDC, which works at the stellar-photometry level), \name\ makes it easier for the model to capture variability that is coming through variations in the centroids and point-spread function from spacecraft pointing, roll, and temperature.

Before we start, a few reminders about the \Kepler\ data are in order:
The spacecraft observed precisely the same field, at fixed pointing (as closely as possible).
About 6 percent of the 96,465,600 pixels in the focal plane are telemetered down to Earth from each 30-min exposure;
  the telemetered pixels are associated with \Kepler\ target stars chosen for study by the \Kepler\ Team (along with a non-trivial set of collateral pixels used for calibration).
The spacecraft is rolled by 90~deg every 93 days (to satisfy Solar-Angle constraints).
The focal plane contains many CCDs;
  this plus the 90-deg rotations means that each star is at a particular location
  on a particular CCD for quarter contiguous periods of time.
The PSF varies strongly across the field and is poorly sampled.
The stars span a huge range in brightness;
  some of \Kepler's most important targets even saturate the device and bleed charge.
The stellar photometry returned by the \Kepler\ SAP and PDC pipelines is based on
  straight, \emph{unweighted} sums of pixels in small patches centered on the stellar centroid.
We will return to this latter point at the end;
  this kind of photometry cannot be optimal;
  it must be possible to do a better job with the photometric measurements.
That's beyond the scope of this project but a place for a valuable intervention on the \Kepler\ data.

We describe the method and deliver all the relevant code in a public, open-source repository.
We also provide an interface to the \Kepler\ data that can be used to produce ``\name\ photometry'' for every \Kepler\ target.

\section{Causal data-driven model}

The first general idea about data-driven modelling of the kind used here
  is that each data point or data source is going to be 
  \emph{predicted} with or using a parameterized mathematical function of \emph{other data}.
That is, given a choice of some \emph{target} \Kepler\ data,
  we are going to find parameters of a function that takes as input some of the other \Kepler\ data,
  and provides as output predictions of the target data.
The simplest models are \emph{linear models},
  in which predictions for the target data are built from linear combinations of the other data,
  and in which the objective functions are quadratic in the prediction residual.
Examples of quadratic objective functions include Gaussian likelihoods, mean-squared-error, and chi-square statistics.
These models are simple,
  not just because they are easy to express and compute,
  but because optimization is convex:
There is only one optimum for the objective function.

The point of these models is to be \emph{flexible},
  so the usual approach is to make the input data set very large,
  and the number of parameters large,
  often as large as---or larger than---the target data available.
Such fits require \emph{regularization} to break degeneracies
  and control ill-constrained parameters.
These regularizations do for optimizations what priors do for posterior probability inferences;
  they express the desired behavior of the fit in the absence of data
  or along directions in parameter space in which the data are not constraining.
Regularizations or priors can break the convexity of linear model fitting;
  if convexity is to be maintained, these regularizations also have to be quadratic,
  or else one of a class of other known forms (one of which is L1-regularization).
Given these considerations, it makes sense to try to build our pixel-level model
  using a linear model with a quadratic objective function and a quadratic regularization;
  this is what \name\ will be.

\begin{figure}[p]
\begin{center}
\includegraphics[width=0.4\textwidth]{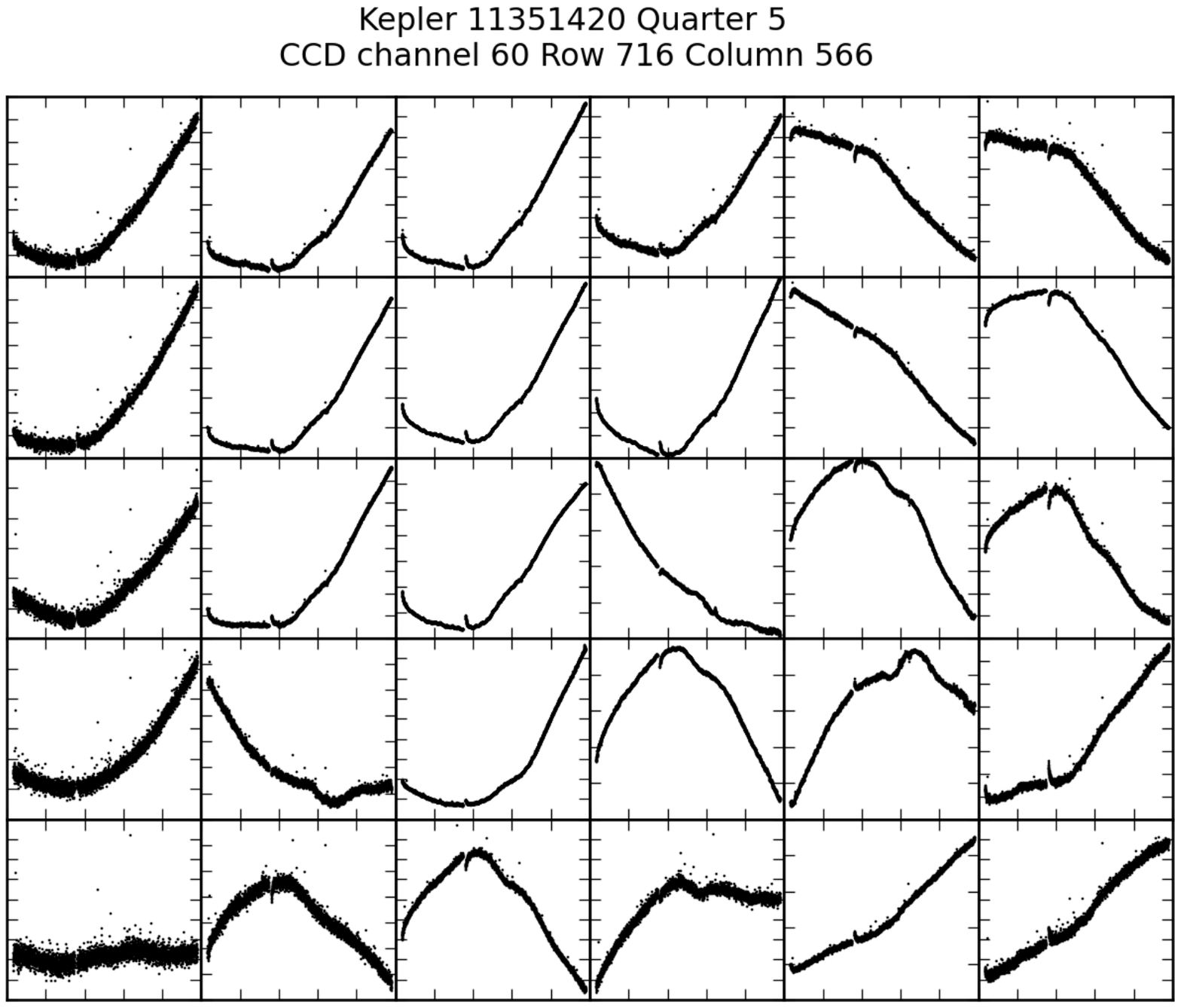}
\includegraphics[width=0.4\textwidth]{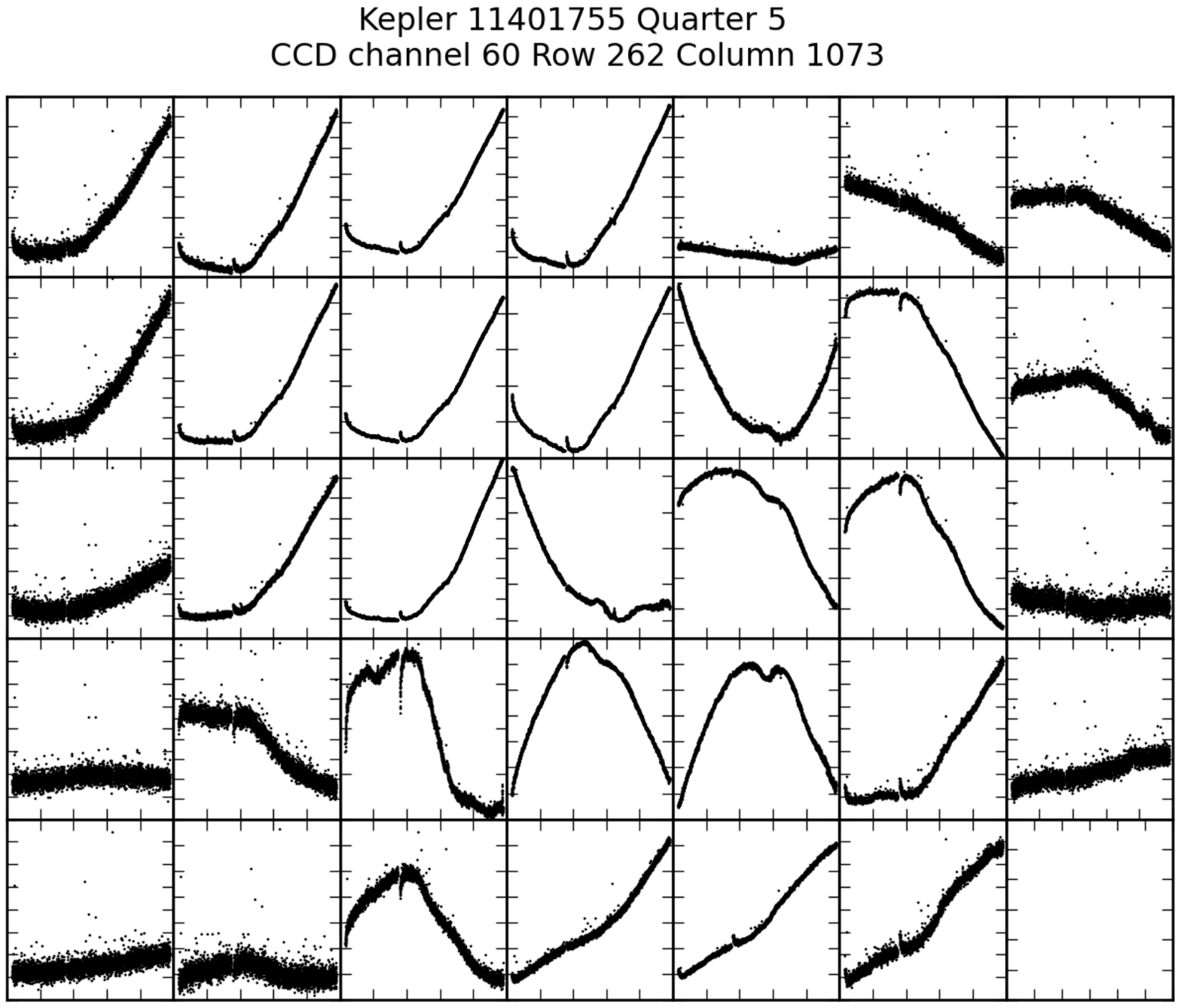}
\end{center}
\caption{
  \label{ccd} 
  Stars on the same CCD share systematic errors. 
  The two panels show pixel fluxes (brightnesses) for two stars in quarter 5. \emph{Left:} KIC 11351420, \emph{Right:} KIC 11401755; 
  here, KIC stands for Kepler Input Catalog. Both stars lie on the same CCD, 
  but far enough apart such that there is no stray light from one affecting the other. 
  Each panel shows the pixels contributing to the respective star.
  Note that there exist similar trends in these two stars, caused by systematic errors. 
}
\end{figure}

The second general idea about data-driven modelling is the investigator's beliefs about the causal processes that generate the data are crucial in restricting the kinds of data that can be used as input to the model.
That is, different assumptions about the physical properties of the data and the data-taking system lead to different structures for the data-driven model.
In the case of \Kepler\ data, if we examine two arbitrary stars far away from each other on the same CCD as in Fig~\ref{ccd} (every grid in the plot is a pixel light curve time series), we can obviously see that the two stars that are hundreds pixel away from each other do have similar trends. 
  If we believe that different stars in the \Kepler\ field vary independently 
  (that is, are not physically synchronized in any way) because of the distance between them, 
  then the only reason that one star might show variability that is strongly \emph{predictive} (useful for prediction) of another star's variability
  is that both stars are being observed by the same device or spacecraft.
That is, one star's pixels can be used to predict another star's pixels inasmuch as spacecraft issues imprint on both stars in related ways;
  they share a common cause---the systematics.  
For another example, if we think the spacecraft is being affected by
  processes that take place over time-scales longer than a single read-out
  (for example, thermal processes),
  then it would be sensible to model the data at time $t$ using not just simultaneous data, but also data from a range of times around $t$.
For another example, if an investigator doesn't care about preserving stellar variability,
  and just wants to detect exoplanet transits (say),
  pixels from the target star can be used to predict pixels from the target star,
  provided they are at large-enough time lags that they don't contain information about
  the signals of interest (i.e., the transits).
That is, if the model is designed to fit not just spacecraft variability
  but also intrinsic stellar variability,
  the predictive model will be permitted to use as input pixels that \emph{do} overlap the target star.
In what follows, we are going to use input data both from pixels of other stars and the target star pixels' past and future.
These models ought to remove both spacecraft variability and intrinsic stellar variability, which is optimized for exoplanet transit searching.

The third general idea is that we need to control for over-fitting.
That is, once you have a flexible-enough model you can in principle fit \emph{anything},
  whether it was caused by the spacecraft, intrinsic stellar variability, or a transiting exoplanet.
How do you prevent a very flexible model from taking small noise-induced fluctuations in all the input data
  and carefully combining them linearly into detailed models for every nook and cranny in the target data?
In most projects in astrophysics, over-fitting is controlled for by limiting model freedom.
The model is restricted in the number of parameters
  (as in ``you can't have more parameters than data points'')
  or by limiting the dimensionality
  (as in principal components analysis)
  or by applying strong priors
  (as with smoothness priors or regularization, 
  that effectively reduce freedom without explicitly reducing the number of parameters).
In each case, the restriction on model freedom is controlled by some \emph{tuning parameters}
  (such as the number of inputs, or number of principal components, or the strength of the smoothness prior).
The tuning parameters can be set or tested with tools like cross-validation,
  the fully marginalized likelihood (Bayes factor or evidence),
  chi-squared statistical criteria,
  or intuition or heuristics.
Here we take a different and more general approach, which is to use a train-and-test framework.

In this framework, the data used to \emph{train} the model%
  ---meaning set the values of the parameters of the model---%
  are disjoint from the \emph{test} data
  ---meaning the data that are being predicted (the target data).
Because we are concerned with detecting exoplanet transits,
  which take a few hours,
  we adopt an extreme version of the train-and-test framework,
  in which the training data are always separated from the test data by many hours.
That is, when we are using the model to predict a particular pixel in the target data taken at time $t$,
  we use parameters obtained by an optimization that makes use of only data
  that comes either at times prior to $t-\Delta t$ or after $t+\Delta t$,
  where $\Delta t$ is a tunable parameter which we will set to $9$~h.
This ensures that no information about any sufficiently short exoplanet transit itself
  can be entering into the prediction of the pixels contributing to the stellar photometry.
In general, if there is a scientific goal of preserving intrinsic stellar variability,
  or transit signals,
  on time-scales of $\tau$,
  the parameter optimization ought to be based on training data taken with a time-exclusion zone of half-width $\Delta t > \tau$.

In addition, 
  each training data set within which we set the parameters (by optimization) has a finite total time extent $t_{\max}\approx 30$\,d.
The train-and-test framework, which includes data out to time $t_{\max}$ but excludes data within $\Delta t$ of the test data,
  effectively assumes that the signals worth preserving have time-scales less than $\Delta t$ 
  but don't recur on time-scales shorter than $t_{\max}$.
Those assumptions are good for our purposes but not necessarily ideal for all users:
Short-period exoplanets and certain kinds of stellar variability
  could be wiped out by a model with these settings of the training-data bounaries.

The fourth general idea involved in this kind of data-driven modelling is that the models often aren't \emph{interpretable}, 
  or at least are hard to interpret.
This means, in particular, that although the model might do a good job of \emph{predicting} the target pixels,
  using a linear (or more complex) combination of the input pixel data,
  it won't deliver anything that can be unambiguously interpreted as the \emph{flux} of the star in question,
  or any other signal we care about.
The data-driven model \emph{effectively} describes the pointing, point-spread function, and flat-field
  of the telescope and camera,
  plus the variability of every star and every exoplanet transit,
  but it does so without ever \emph{explicitly} creating any of those objects.
We have to make some kind of heuristic or interpretive move to extract from the data-driven model the quantities of interest.

Finally, the fifth general idea is that there is no objective \emph{ground truth} against which we can tell
  that any particular data-driven model is better or worse than any other.
This problem is a problem for \name, and for the \Kepler\ PDC, and any other data-driven calibration models.
One might think that the ``best'' model is the one that predicts data with lowest variance;
  this would be true if all models adhered to the same train-and-test framework, which they don't.
Besides, a model that can predict not just the spacecraft-induced variability,
  but also variability caused by exoplanet transits of interest, will effectively over-fit and distort the most important information in the light curves.
That is, what is considered best for modelling the data depends on the objectives of the user.
For us, who are interested (in the long term) in finding and characterizing Earth analogs,
  the ``best'' data-driven model is the one that produces the most success in finding and characterizing Earth analogs!
What we will show in what follows is that the \name\ photometry does not distort
  artificial exoplanet transit signals injected into real \Kepler\ pixel-level data.
In the end, the value of the \name\ will be demonstrated by the scientific projects it enables.

We also want to emphasis that all the calibration in this paper for \Kepler\ is based on the assumption that the device can be self-calibrated, which means all the information we need to calibrate the device are contained in the science data and no external information, for example, measurements from any other device, are needed.
Elaboration of the method in a machine learning context can be found in \citealt{icml2015}.

\section{\name\ specification and tuning parameters}
\subsection{\name\ model}
Each individual \Kepler\ target star is, for each quarter, in a particular location on a particular CCD on the \Kepler\ focal plane.
Associated with each target star is some set of pixels, located in a small contiguous patch centered on the target star, and telemetered down once for every 30-min exposure.
Here we are ignoring all short-cadence data that are taken in a different mode with shorter exposures.
That is, each telemetered pixel in each CCD is associated with a particular \Kepler\ target star.

Consider a particular pixel $m$ in the focal plane in one particular month
  (here we always use data per month rather than quarter,  since there are discontinuities between every months).
In that month, the spacecraft delivers $N\sim 1300$ measurements $I_{m,n}$
  of the intensity falling in pixel $m$ at the $N$ times $t_n$ at which exposures were taken during the month.
(For intensity measurements $I_{m,n}$ we are using pixels from the FLUX column of Target Pixel File, which is flux of each pixel after processed by the pipeline module CAL, the removal of the interpolated background, and the removal of cosmic rays on MAST
  \footnote{\url{http://archive.stsci.edu/kepler/}}, for details of the TPF, 
  see the \project{Kepler Archive Manual} 
  \footnote{\url{http://archive.stsci.edu/kepler/manuals/archive_manual.pdf}}).
We want to build a prediction for the intensity $I_{m,n}$ in pixel $m$ at each time $t_n$
  using the intensities $I_{m',n}$ of other pixels $m'$.
The question is:  What other pixels to choose?
There are many possible qualitatively and quantitatively different choices here.
Assuming that they are affected by similar systematic errors as the target star pixels, 
  we choose all the pixels $m'$ associated with the $Q$ stars on the same CCD in the same quarter that are closest in magnitude
  (\Kepler\ magnitude as reported in the \Kepler\ Input Catalog)
  to the target star associated with pixel $m$.
This set of pixels%
  ---from the $Q$ stars on the same CCD---%
  is the \emph{pixel set} $\set{M}_m$ associated with pixel $m$.
To minimize the effects of blending with, and crosstalk from, nearby stars, we require that the predictor pixels are from stars at least 20 pixels away from the target star.
Note that because the set $\set{M}_m$ is of pixels associated with \emph{different} stars
  than the star associated with pixel $m$, and are far from it on CCD, 
  the pixel $m$ will not be in the set $\set{M}_m$,
  and nor will any of pixel $m$'s close neighbors.
That is, there will be no (or almost no) overlap in stellar illumination of pixel $m$
  and the pixels $m'\in\set{M}_m$.

\begin{figure}[p]
\begin{center}
\includegraphics[width=0.8\textwidth]{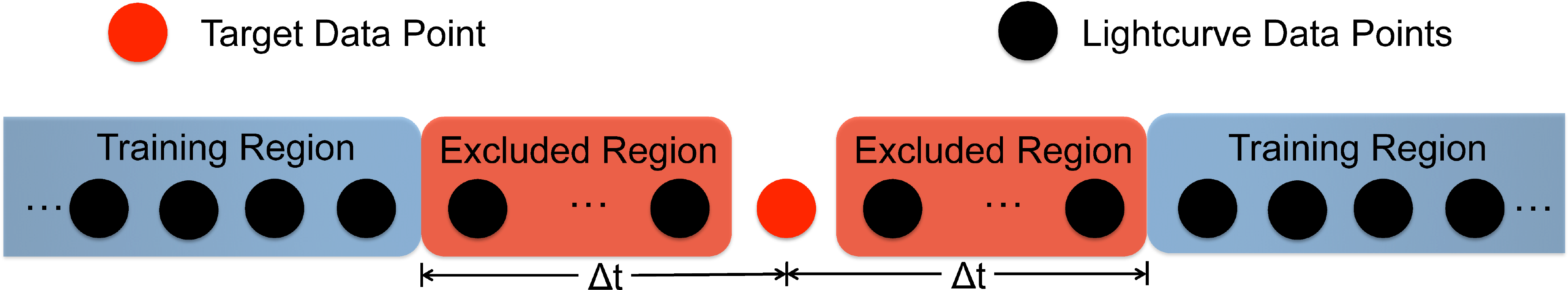}
\end{center}
\caption{
  \label{train-and-test} 
  Train-and-test framework. 
  Data near the target point $t_{n}$ within $\Delta t$ is excluded in the training set}
\end{figure}

When the measurement $I_{m,n}$ is being predicted from pixel $m$ at a particular time $t_n$,
  a train-and-test framework is used, in which not only
  data at time $t_n$ are not used in optimizing the parameters of the regression or fit,
  but we don't even use any times $t$ such that $|t-t_n| < \Delta t$,
  where $\Delta t=9$\,h as shown in Fig.~\ref{train-and-test}.
That is, the model is trained (are optimized) by using the \emph{time set} $\set{N}_n$ of time
  indices $n'$ such that for all $n'\in\set{N}_n$,
  $t_{n'}$ is in the same month as $t_n$,
  and $|t_{n'} - t_n|>\Delta t$.
The time set of indices $\set{N}_n$ therefore does not overlap the time point $t_n$,
  nor any of its neighbors in a time window of half-width $\Delta t$. 
The width of the window $\Delta t$ is set to be at least the duration of the transits to preserve transits while still predicting the systematics and stellar variability well.
  
In addition to the pixel set $m'\in\set{M}_m$ from other stars,  
  we also include the past and future of the target star, i.e., an autoregressive (AR) component, as input 
  to remove more of the stellar variability and thus increase the sensitivity for transits. 
To do this, we select an exclusion window of half-width $\Delta t$ including R data points, 
  where $\Delta t=9$\, h (as big as the window in the train-and-test framework) 
  around the point of time $t_{n}$ being corrected, 
  to ensure that we do not remove the transit itself 
  and then use $S$ closest future and past time points subject to that exclusion. 
That is, for every pixel $m$ in the target star we construct $S$ virtual time series, 
  in which every component $v$ is defined to be     
\begin{eqnarray}
I_{v,n} = I_{m,n-R-k}\ or\ I_{v,n} = I_{m,n+R+k}
\quad,
\end{eqnarray}
where $k = 1, 2\dots, S,$ so if there are totally $M$ pixels in the target stars, 
  we will finally add $2\cdot M\cdot S$
  autoregressive components into the predictors, which constructs the autoregressive set $\set{V}_m$.

As mentioned above, in the context of this project, a model is something that can predict data given settings of some parameters, and an objective function that can be used to find best values for those parameters.
The \name\ treats each \Kepler\ data point $I_{m,n}$ as being predictable from a linear combination of data points $I_{m',n}$, where $m'$ is from the set of non-overlapping pixels $m'\in\set{M}_m$.
\begin{eqnarray}
I_{m,n}         &=& I^{\ast}_{m,n} + e_{m,n}
\\
I^{\ast}_{m,n}  &=& \sum_{m'\in\set{M}_m} a_{m,n,m'}\,I_{m',n} + \sum_{v\in\set{V}_m} b_{m,n,v}\,I_{v,n}
\quad,
\end{eqnarray}
where $I^{\ast}_{m,n}$ is the prediction (by the model) for data point $I_{m,n}$,
  $e_{m,n}$ is a noise contribution or residual away from the prediction,
  and the $a_{m,n,m'}$ and $b_{m,n,k}$ are the parameters (linear coefficients of the prediction).
The parameters $a_{m,n,m'}$ and $b_{m,n,k}$ have indices $m$ because
  they are different for every pixel $m$,
  and they will even be different for every time step $n$;
  that is, there will be a separate best-fit value for the parameters for every $I_{m,n}$.
And the pixel $I_{m',n}$ from other stars at the same time point n is used as input to make prediction.
If we presume that the residuals away from the prediction are normally distributed with zero mean and known variance, then likelihood optimization reduces to $\chi^2$ minimization.
We add to the standard chi-squared definition a regularization term
  (equivalent to multiplying the likelihood by a Gaussian prior pdf)
  that penalizes large absolute values for the the coefficients $a_{m,n,m'}$:
\begin{eqnarray}
\chi^2_{m,n}    &=& \sum_{n'\in\set{N}_n} \frac{[I_{m,n'} - I^{\ast\ast}_{m,n',n}]^2}{\sigma^2_{m,n'}}
                 + \lambda_{a}\sum_{m'\in\set{M}_m}a_{m,n,m'}^2 + \lambda_{b}\sum_{v\in\set{V}_m}b_{m,n,v}^2
\\
I^{\ast\ast}_{m,n',n} &=& \sum_{m'\in\set{M}_m} a_{m,n,m'}\,I_{m',n'} + \sum_{v\in\set{V}_m} b_{m,n,v}\,I_{v,n'}
\quad,
\end{eqnarray}

where the $\sigma^2_{m,n'}$ are the (the column FLUX\_ERR in TPF is used, which is a image of 1-sigma error provided by the Kepler team) individual-pixel noise variances, and $\lambda_{a}$, 
  $\lambda_{b}$ set the strength of the regularization (or width of the prior pdf) for parameters $a_{m,n,m'}, b_{m,n,v}$. 
  In general, $\lambda_a$ and $\lambda_b$ could depend on $m,n$, however, we do not make use of this freedom. 
  Here the pixel value $I^{\ast\ast}_{m,n',n}$ is different from the prediction $I^{\ast}_{m,n}$. It is not the prediction for time $n'$ using $a_{m,n',m'}, b_{m,n',v}$ as coefficient, but using the coefficient $a_{m,n,m'}, b_{m,n,v}$ that is trained for prediction in time n to set the value. It is purely an intermediate product in the process of training the model. It has three indices, 
  because in the \name\ the $\chi^2$ minimization for every time
  point is independent. Details of the independent $\chi^2$ minimization will be discussed below.

The train-and-test idea comes into the objective function $\chi^2_{m,n}$:
When we are computing the objective function $\chi^2_{m,n}$,
  we are using only time points $n'\in\set{N}_n$ that don't overlap the target time point $n$.
We train the model---that is, set the parameters $a_{m,n,m'}$ and $b_{m,n,v}$---%
  by choosing the full set of parameters $a_{m,n,m'}$ and $b_{m,n,v}$ 
  that jointly minimize the objective function $\chi^2_{m,n}$.
Fortunately, given the form of the model,
  this minimization is just a linear solve.
Importantly however, and perhaps surprisingly, the objective function $\chi^2_{m,n}$ is \emph{different}
  for every target data point (pixel value to be predicted) $I_{m,n}$;
  we have to do an \emph{independent optimization}
  of the parameters for every pixel value we want to predict at every time.
That is, every pixel datum $I_{m,n}$ we predict will have
  \emph{different} settings of the parameters $a_{m,n,m'}$ and $b_{m,n,v}$.
Since there are of order $10^{6}$ total pixel values in the \Kepler\ Target Pixel Files for \emph{each \Kepler\ target},
  and of order $10^{3}$ data sources used as basis functions in the linear fitting,
  this represents a lot of linear solves, and of order $10^{9}$ parameters to obtain and record.

Once we have obtained the parameters $a_{m,n,m'}$ and $b_{m,n,v}$ corresponding to a particular data point $I_{m,n}$,
  we can make a \emph{prediction} $I^{\ast}_{m,n}$ for the intensity at that pixel.
By construction of our objective function, this is truly a prediction,
  in the sense that the optimization that produced the parameters
  did not make use of $I_{m,n}$ itself,
  nor even any of the pixel values that are nearby in angle or time.
  With all the predicted pixel values $I^{\ast}_{m,n}$ of the target star,  
  our new stellar flux estimates,
  what we will call the ``\name\ prediction'',
  is constructed from the \name\ pixels
  in precisely the same way as the official \Kepler\ SAP photometry
  is constructed from the Target Pixel Files.
That is, we perform a weighted sum of \name\ predicted pixels $I^{\ast}_{m,n}$ with weights $w_m$,
  all of which are either zero or unity,
  and we adopt precisely the same weight assignments as are adopted in the SAP photometry;
  in equations, the flux estimate $S_n$ at time $t_n$ of the target star is given by
\begin{eqnarray}
S_n = \sum_m w_m\,I^{\ast}_{m,n}
\quad ,
\end{eqnarray}
where the sum is over the $M$ pixels that are associated with the target star and the weight $w_m$ is unity for pixels in the optimal
aperture while zero outside the optimal aperture.
As we have noted above and below, these photometric estimators are not optimal for any purpose, and chances are that the aperture may not be able to capture all the light from the star or there can be issues of crowding and consequent contamination flux from nearby stars, but improving them is beyond the scope of this project.

Because of the train-and-test framework,  we expect the \name\ prediction $S_{n}$ to be a prediction of what the star light curve would have under the influence of the spacecraft but with no exoplanet signal. 
Based on this idea, 
  we can construct the ``\name\ flux'' 
  (where the inverted commas indicate that it is not a flux in the sense that stellar variability is not preserved) 
  to be the relative residual between \name\ prediction and SAP Flux $F_{n}$
\begin{eqnarray}
\delta_{n}&\equiv&\frac{F_{n} - S_{n}}{S_{n}}
\quad .
\end{eqnarray} 
In some sense, it is this relative residual---the \name\ flux---that contains the exoplanet transit signals we seek. 
That is, a transit creates a negative residual away from the prediction, 
  and the amount of relative residual is just the fraction of the light eclipsed by the planet (the transit depth). 

\subsection{Tuning parameters}
\begin{figure}[p]
\begin{center}
\includegraphics[width=\textwidth]{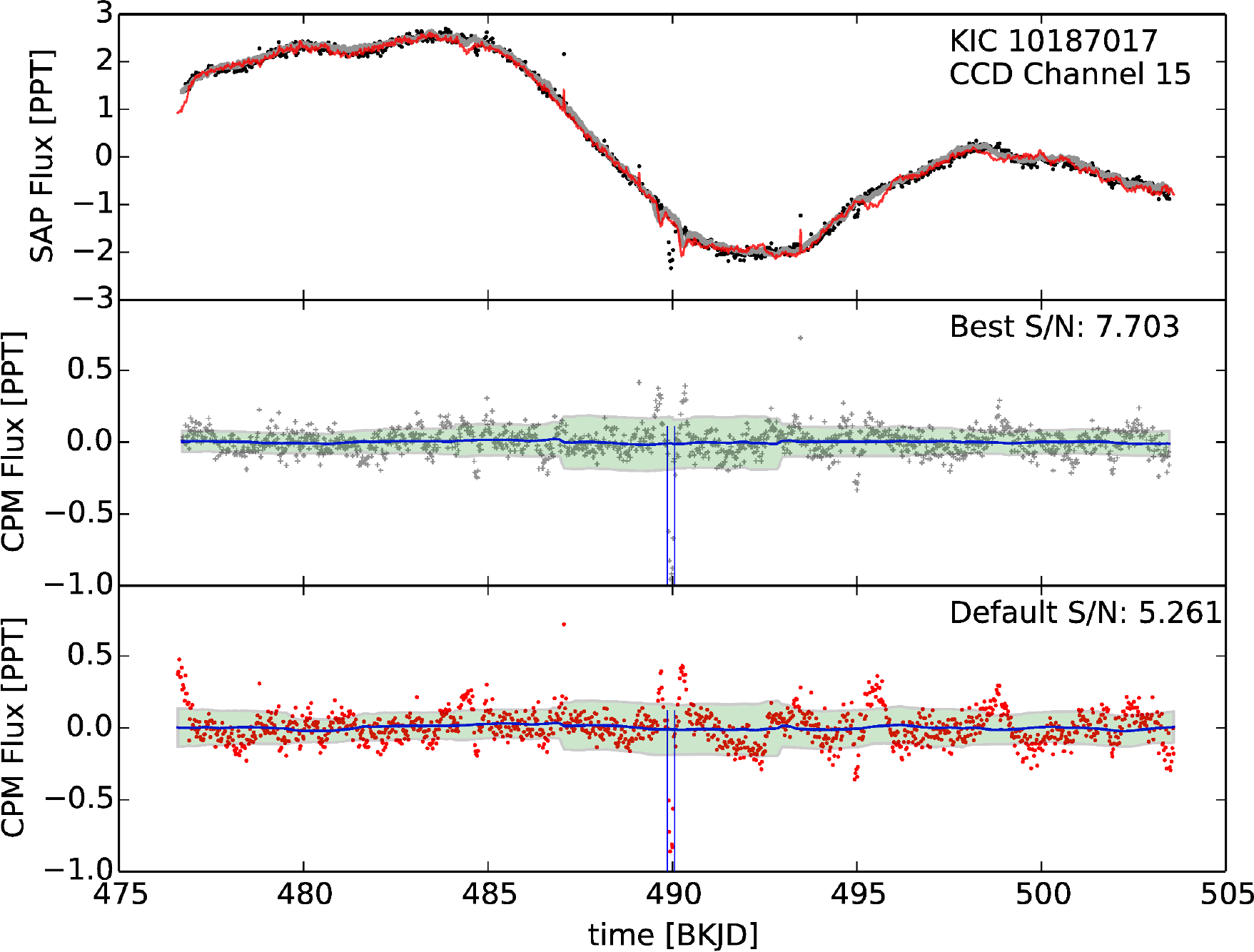}
\end{center}
\caption{
  \label{hyperparameter}
  Comparison between default and optimized tuning parameters. 
  In the top panel, SAP flux is plotted in black, 
    \name\ prediction of default tuning parameters is in thin red line, 
    \name\ prediction of optimized tuning parameters is in bold grey line. 
  The middle panel shows the \name\ flux of the optimized tuning parameters, 
    while \name\ flux of the default tuning parameters is in the bottom panel, 
    the two vertical blue lines indicate the location of the injected transit signal, 
    and the signal to noise ratio is calculated in both situations. 
  The optimized tuning parameters perform better than the default, 
    but the results from both situations are similar.
  Here the signal strength is determined by the mean depth of the transit signal,
    while the noise level is estimated with the root-mean-square deviation of the points in a 6-day running window.}
\end{figure}

The \name\ has 4 tuning parameters---the number of predictor stars Q (or number of predictor pixels P), the number of autoregressive components S, and the two regularization strengths $\lambda_{a}$ and $\lambda_{b}$.
To set these 4 tuning parameters, in principle we need to run something like a cross-validation on every single star to optimize performance.
But optimizing in this 4-dimensional space is expensive, 
  especially since in the \name\ we need to solve thousands of linear systems, 
  each of which has thousands of free parameters. 
Therefore, in this paper, 
  we just set a general set of ``default'' tuning parameters ($P=4000$, $S=3$, $\lambda_a=1e5$, $\lambda_a=1e5$), 
  which works well in most of the stars without high variability. 
To show how this general set of tuning parameters performs well, 
  we present a few comparisons between optimized and default tuning parameters in Fig~\ref{hyperparameter}.
We can see that in Fig~\ref{hyperparameter}, 
  there is less variation in the \name\ flux 
  when we use optimized tuning parameters (grey points in second panel, $Q=60$, $S=3$, $\lambda_a=1e7$, $\lambda_a=1e5$) 
  than when we use the default values (red points in the bottom panel). 
In addition, the optimized flux has higher signal-to-noise ratio on the known transit than the default. 
However, the performance of default tuning parameters is acceptable 
  and we will use this set of tuning parameters throughout the rest of the paper 
  to show the general performance of \name. 
It is also worth mentioning that the tuning parameters are optimized only on a relatively coarse grid, which means the performance can still be improved when setting the tuning parameters more precisely.
There are also parameters such as train-and-test window size, the CCD constraint on the predictor stars, the minimum distance of the predictor stars and the ranking algorithm to select predictor stars. In \name\ we set these parameters heuristically, but they could be open for discussion.
  
\section{Examples and results}

\subsection{Effect on transit signals}
\begin{figure}[p]
\begin{center}
\includegraphics[width=0.325\textwidth]{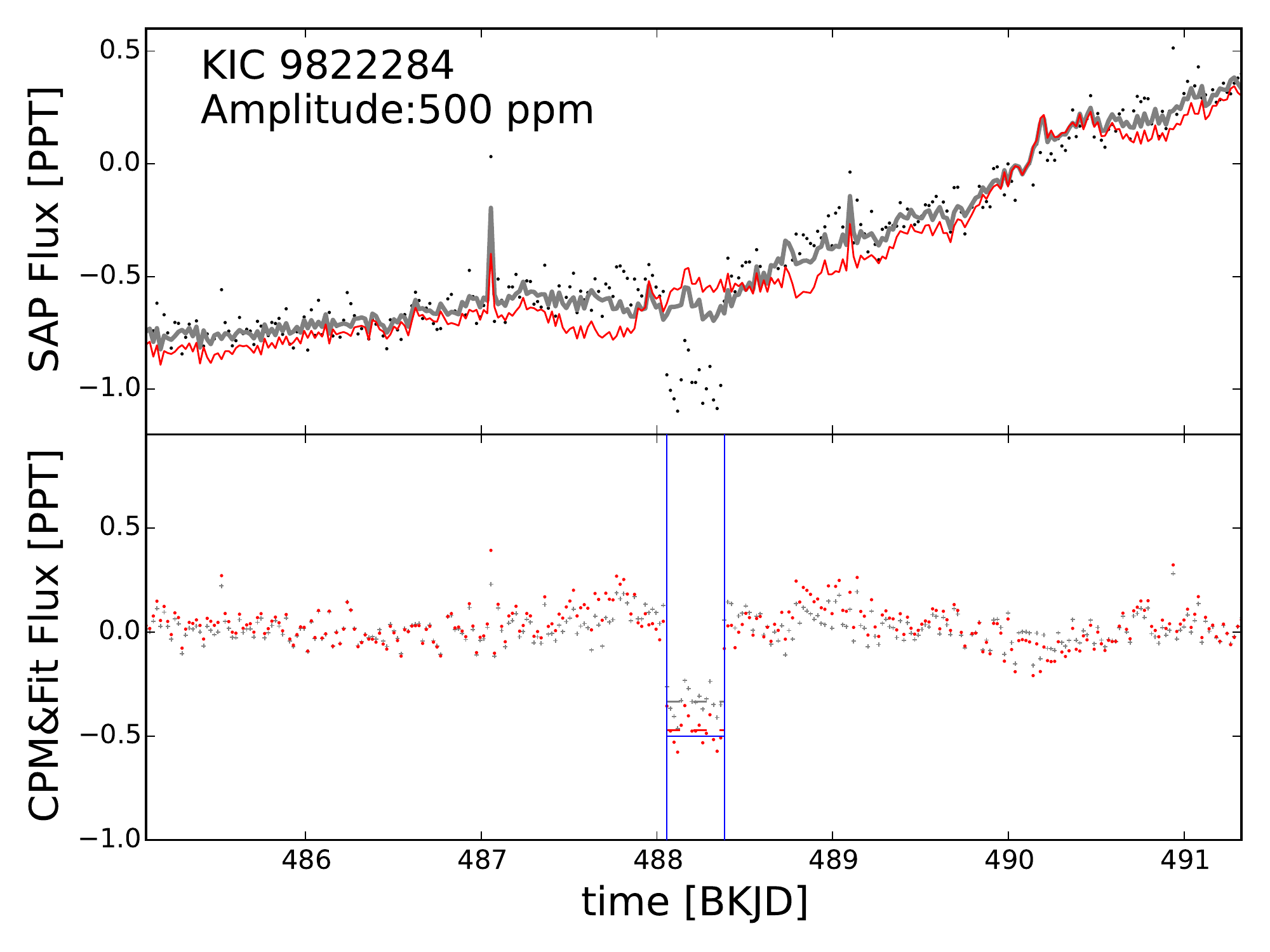}
\includegraphics[width=0.325\textwidth]{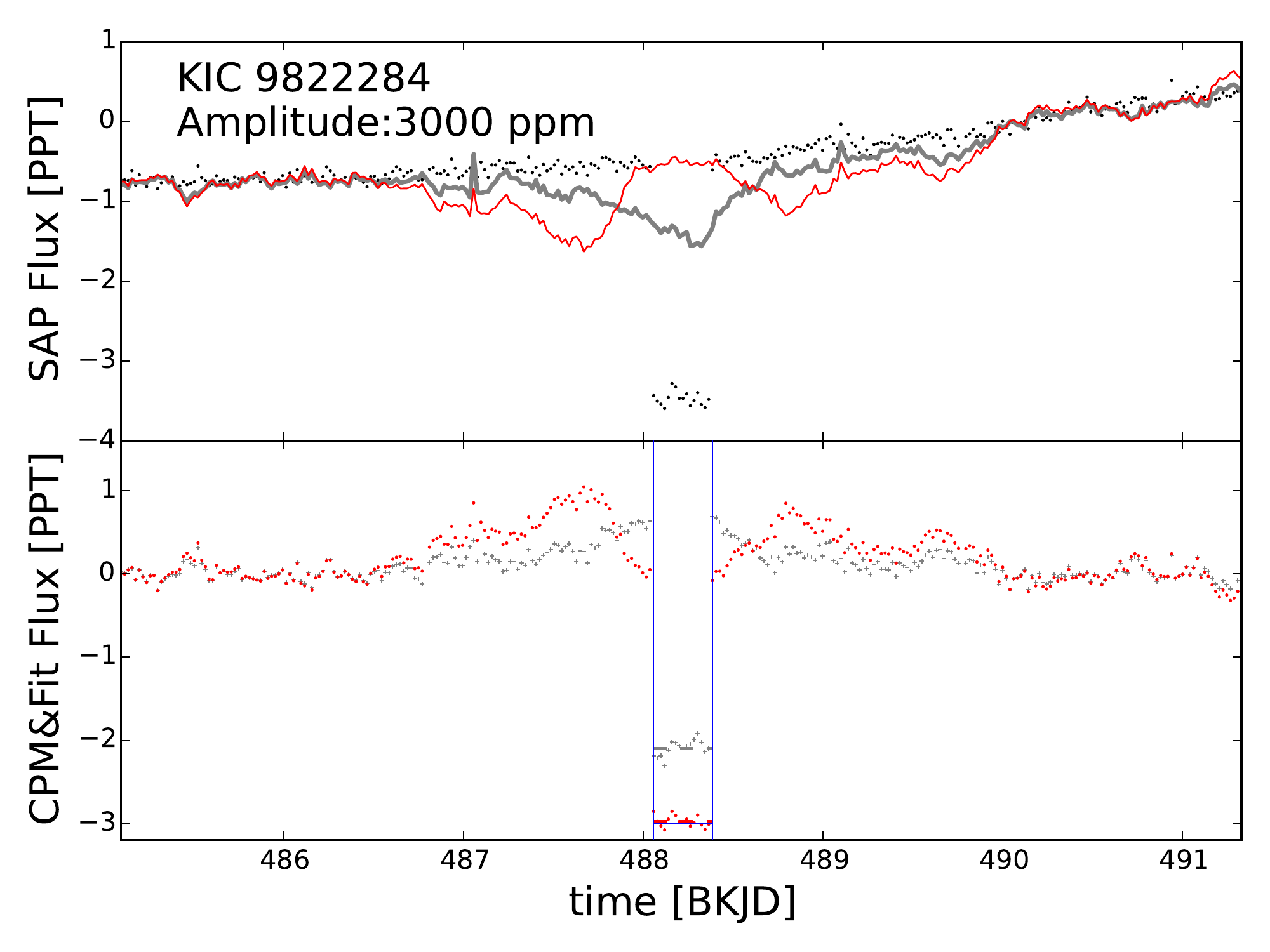}
\includegraphics[width=0.325\textwidth]{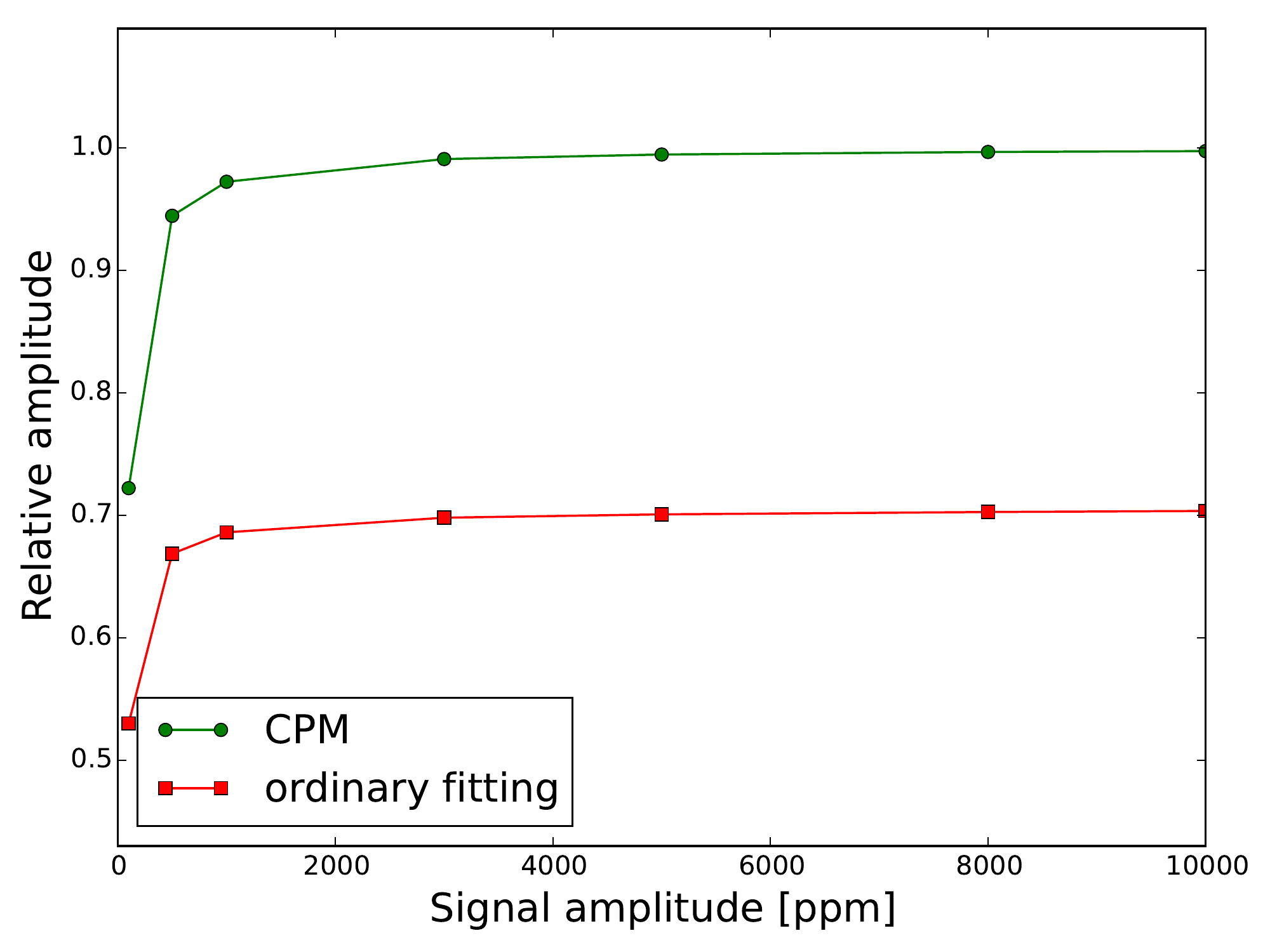}
\end{center}
\caption{
  \label{distortion} 
  \name\ preserves transit signals.
  Simple box models with amplitude from 100 ppm to 1 percent were injected into the light curve of star KIC 9822284. 
  \emph{Left:} An example of light curve with 500 ppm signal injected. In the top panel, SAP flux is plotted in black dots, the \name\ prediction is in thin red line, ``ordinary fitting'' is in bold grey line.
  In the bottom panel, red points and grey cross represent the \name\ flux and the ``ordinary fitting'' flux, while the signal amplitude is indicated by the horizontal line. 
  \name\ preserves the original signal amplitude while the ``ordinary fitting'' overfits the signal. But the \name\ flux is distorted a little around the edge of the box model.
  \emph{Middle:} The same plot for light curve with 3000 ppm signal injected. There is more distortion with a stronger signal.
  \emph{Right:} The relative signal amplitude measured in \name\ and ``ordinary fitting'' flux is plotted as a function of the injected signal amplitude. The \name\ can consistently preserve more signal than the ``ordinary fitting".}
\end{figure}

One important feature of \name\ is that we use the train-and-test framework to preserve the transit signal. 
To show how well the train-and-test framework performs, we perform the following experiments on \Kepler\ data shown in Fig~\ref{distortion}:
In the light curve of KIC 9822284, simple rectangular box model with different amplitude (from 100 ppm to 1\%) is injected. 
To do the injection, each pixel light curve of the target star is multiplied by a factor of the amplitude within a timescale of 8 hours. 
With the distorted light curve, both \name\ and ``ordinary fitting" are applied to the data. 
By ``ordinary fitting", we mean fitting with all time points,  in comparison with the \name\, in which for each data point the prediction is made without the data in the excluded region. 
Both methods fit the light curve quite well outside the range of the inserted signal. 
However, the ``ordinary fitting'' fits out a large portion of the signal, while the \name\ preserves the original amplitude. 
This simple experiment shows that the train-and-test framework is capable of preserving the transit signal. 
This is crucial for searching for exoplanets, especially Earth-like planets, which only have $\backsim$ 100\ ppm signals. 
But there are issues: 
The \name\ does not predict the light curve very well around the box model, since in these regions the prediction is under the strong influence of the transit signal. 
Thus the \name\ will introduce some distortion around the transit signal, and we expect that the distortion increases with the depth of the signal. This issue will be discussed in detail in Section 5.

\begin{figure}[p]
\begin{center}
\includegraphics[width=0.32\textwidth]{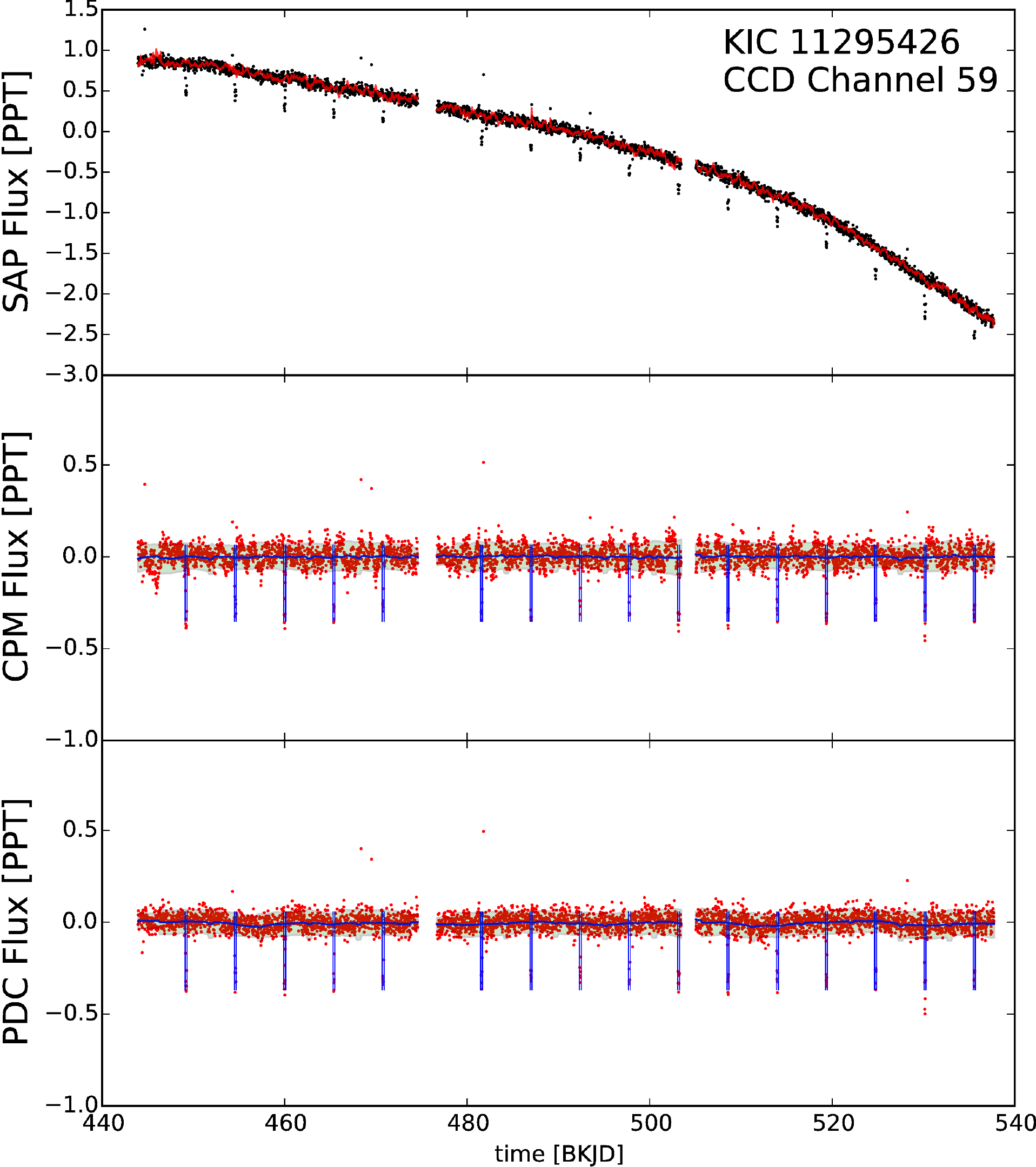}
\hfill
\includegraphics[width=0.32\textwidth]{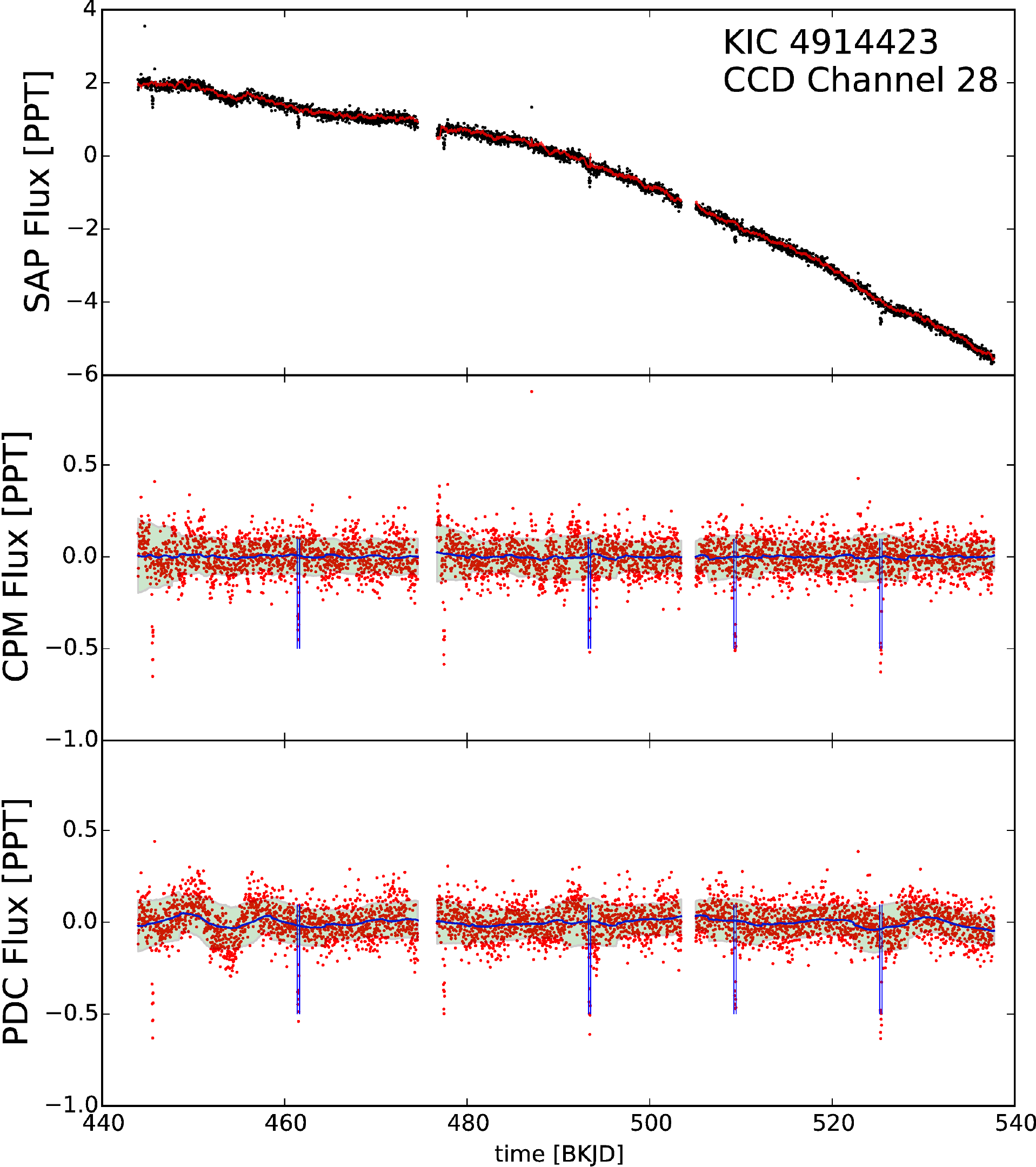}
\hfill
\includegraphics[width=0.32\textwidth]{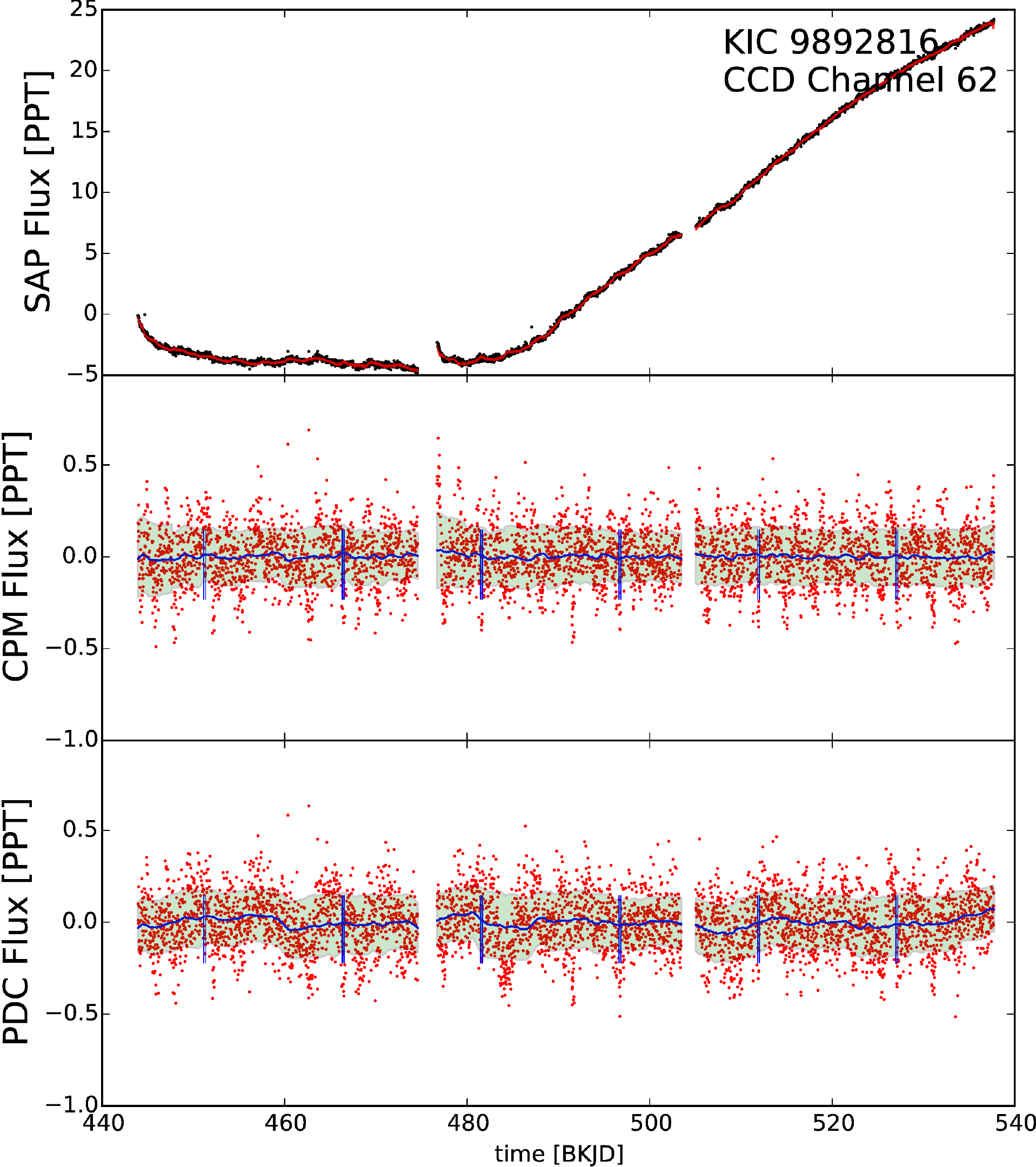}

\includegraphics[width=0.32\textwidth]{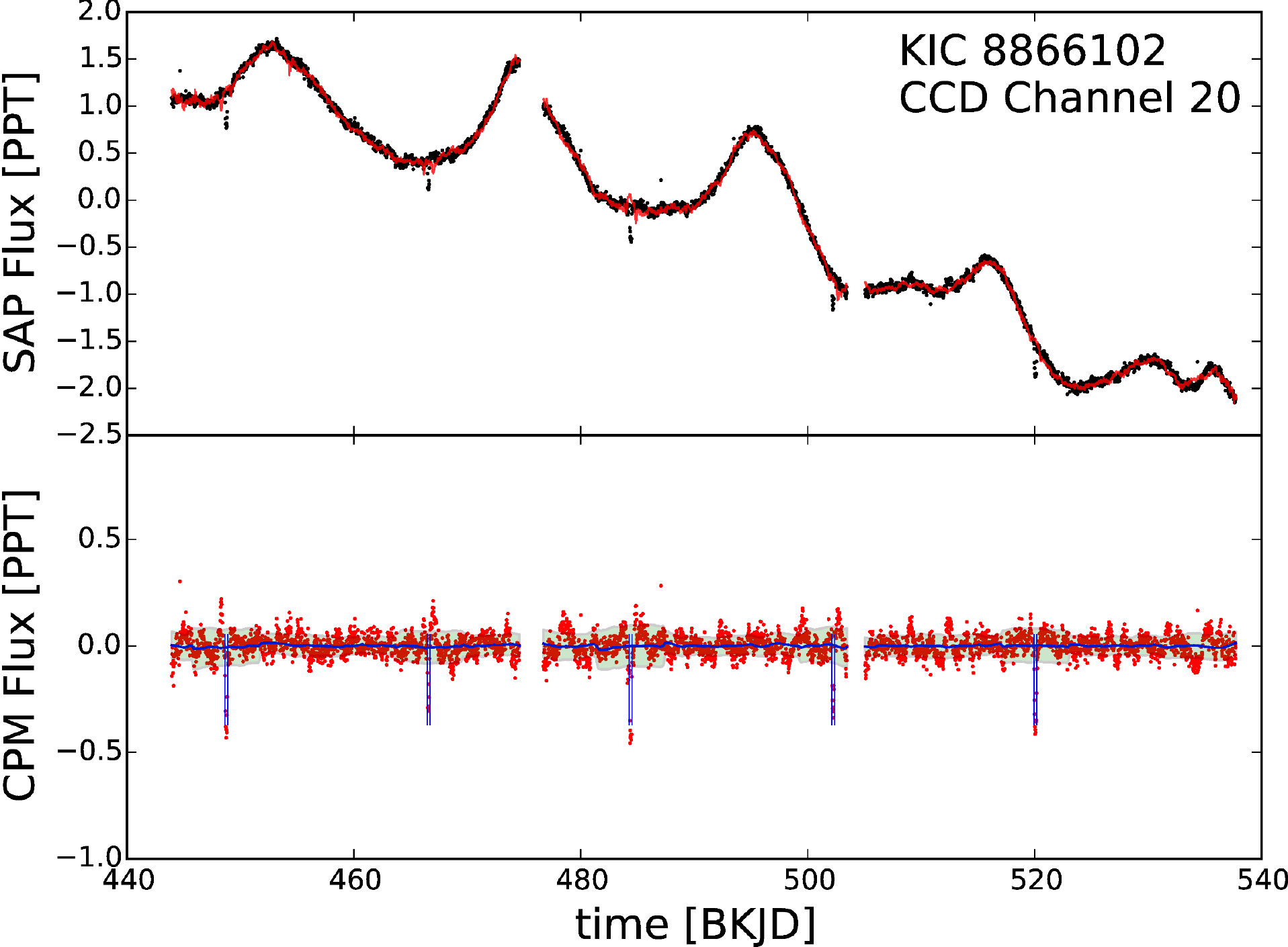}
\hfill
\includegraphics[width=0.32\textwidth]{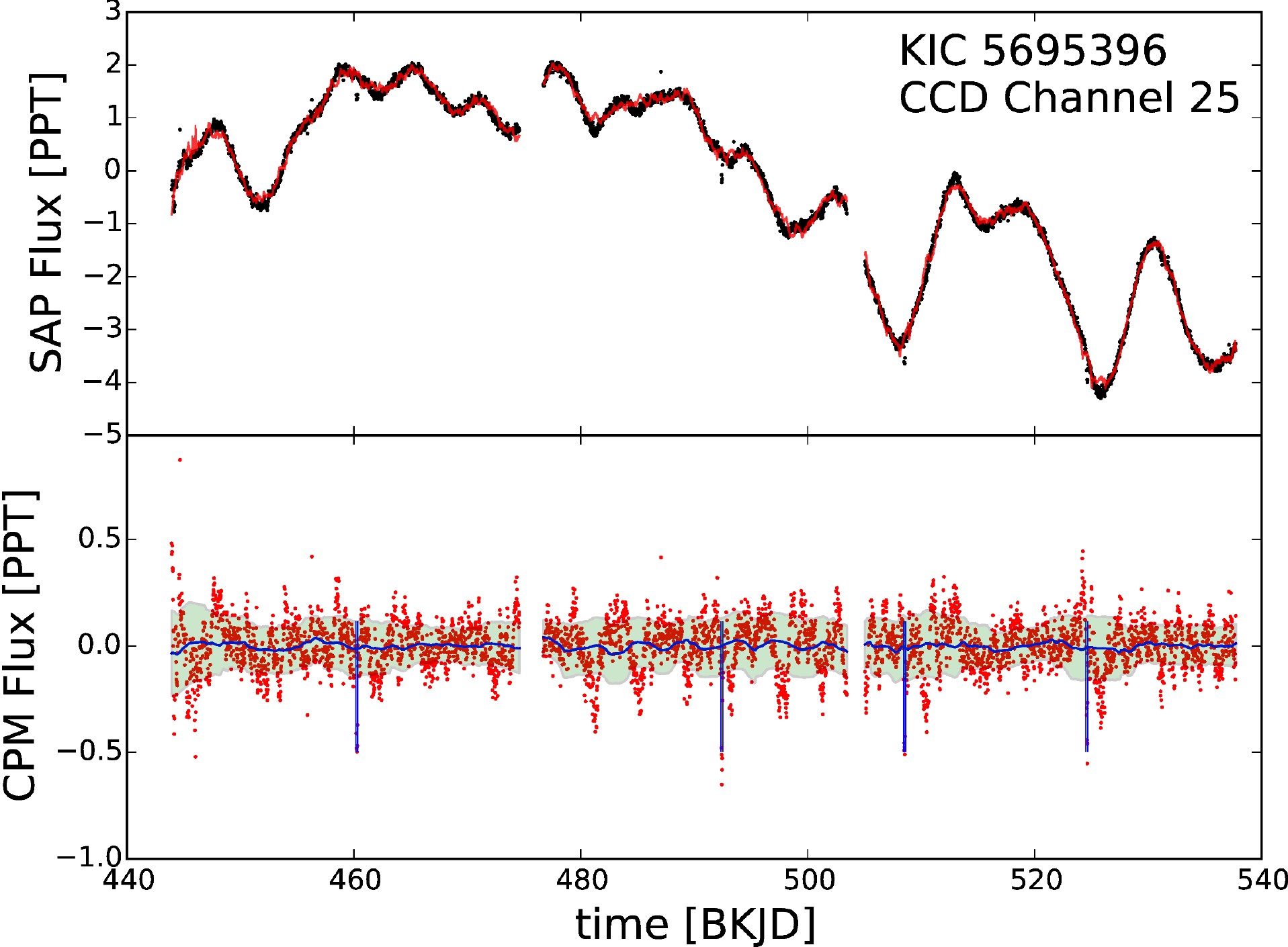}
\hfill
\includegraphics[width=0.32\textwidth]{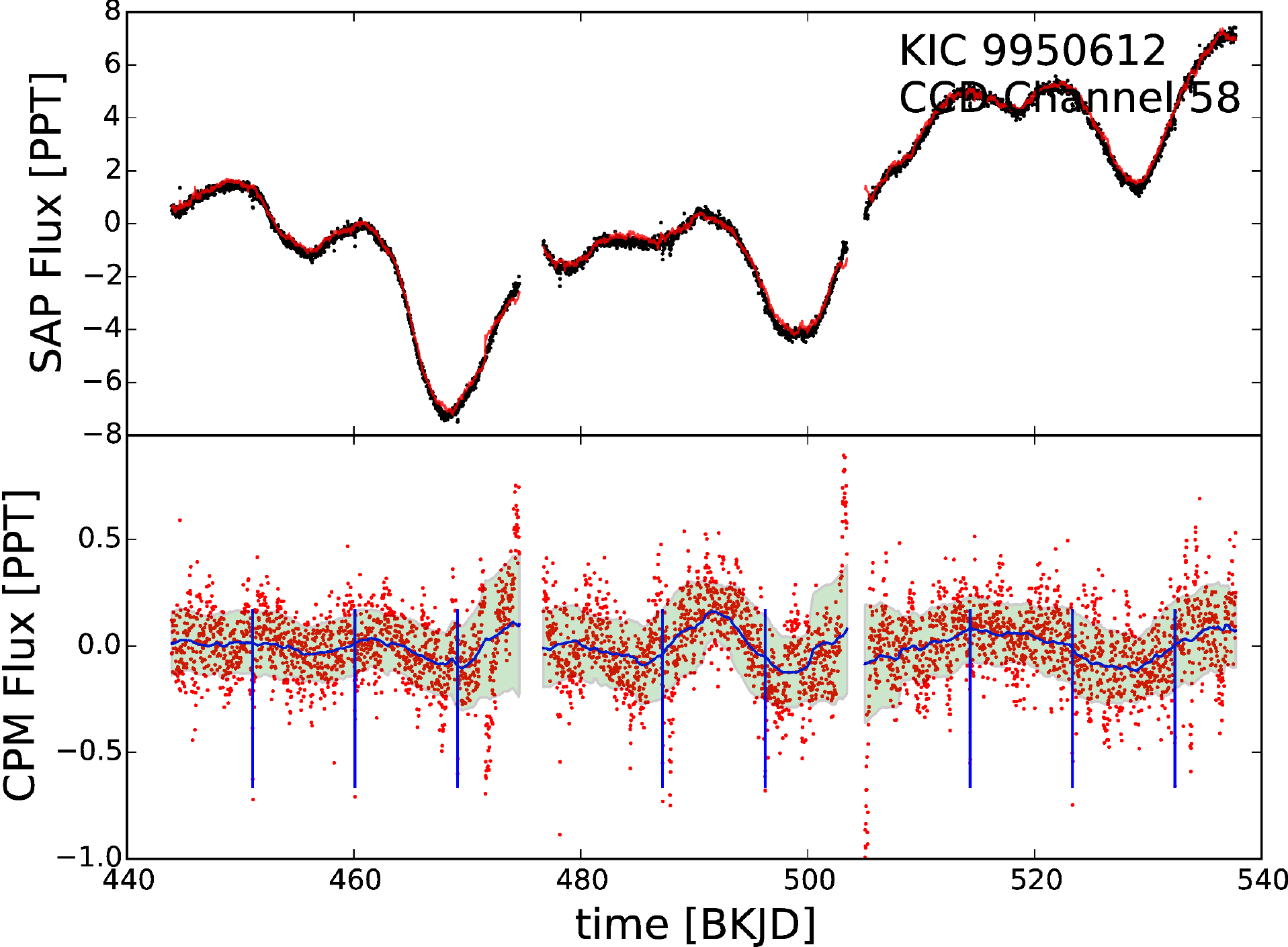}

\includegraphics[width=0.32\textwidth]{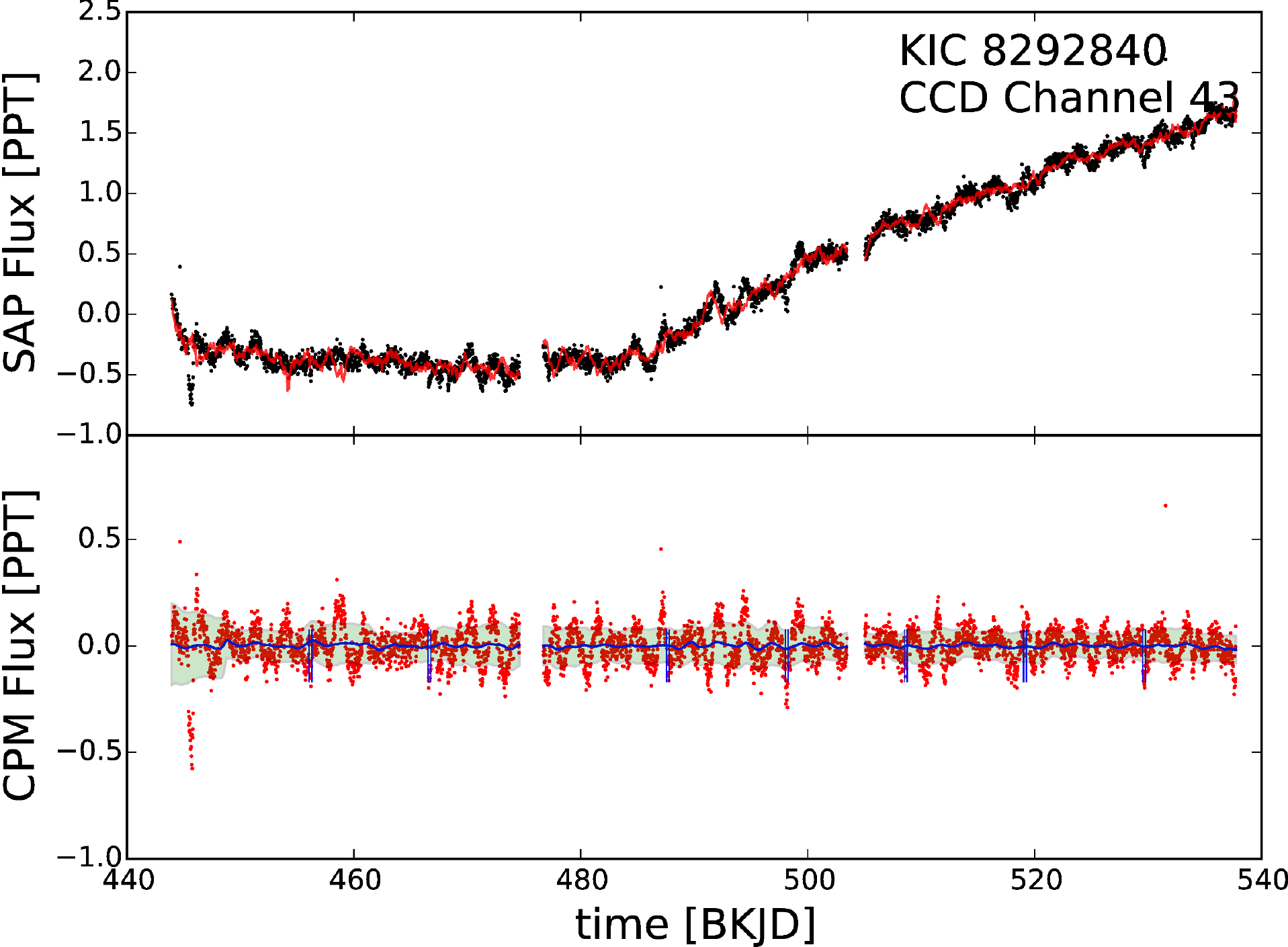}
\hfill
\includegraphics[width=0.32\textwidth]{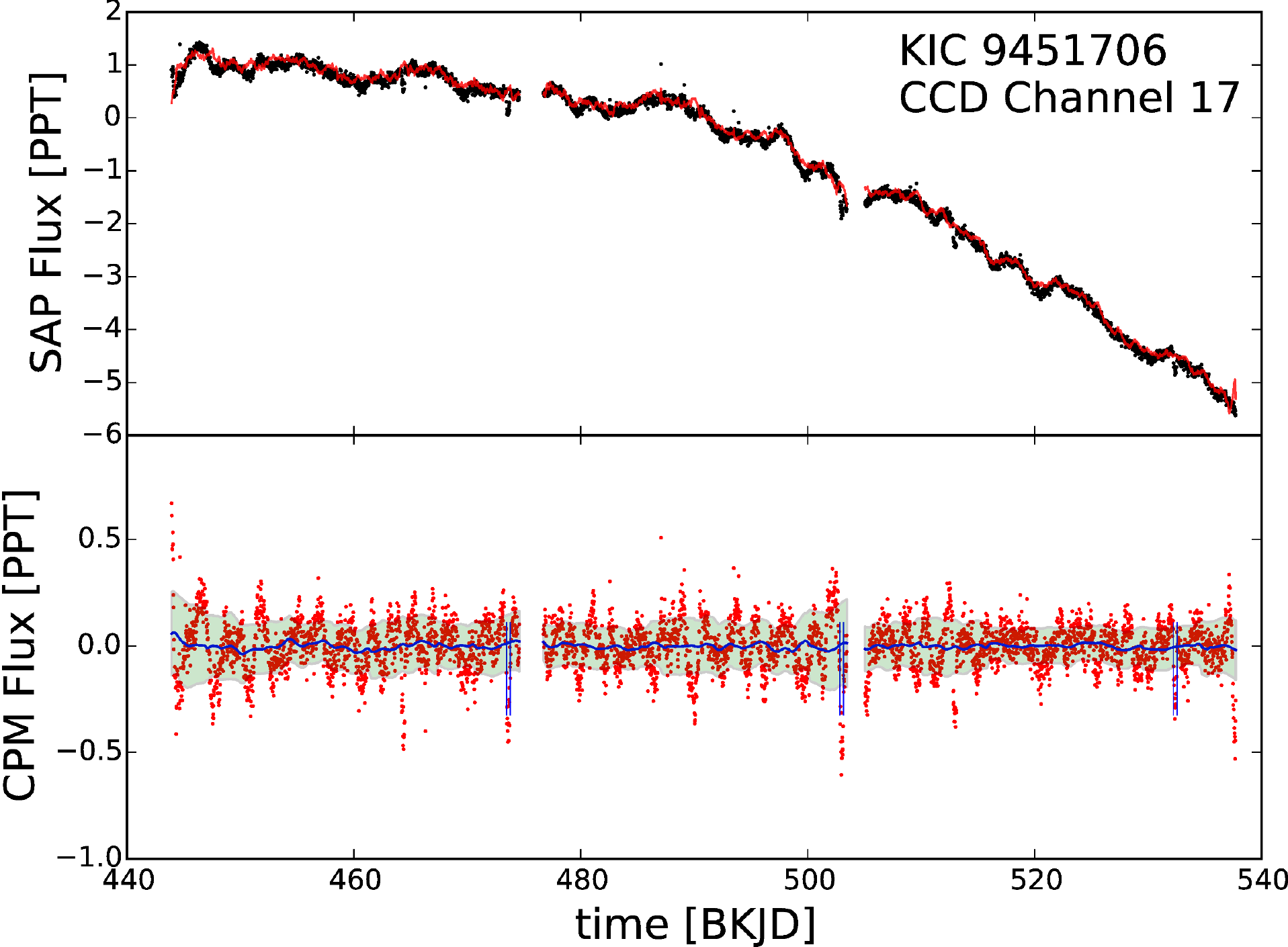}
\hfill
\includegraphics[width=0.32\textwidth]{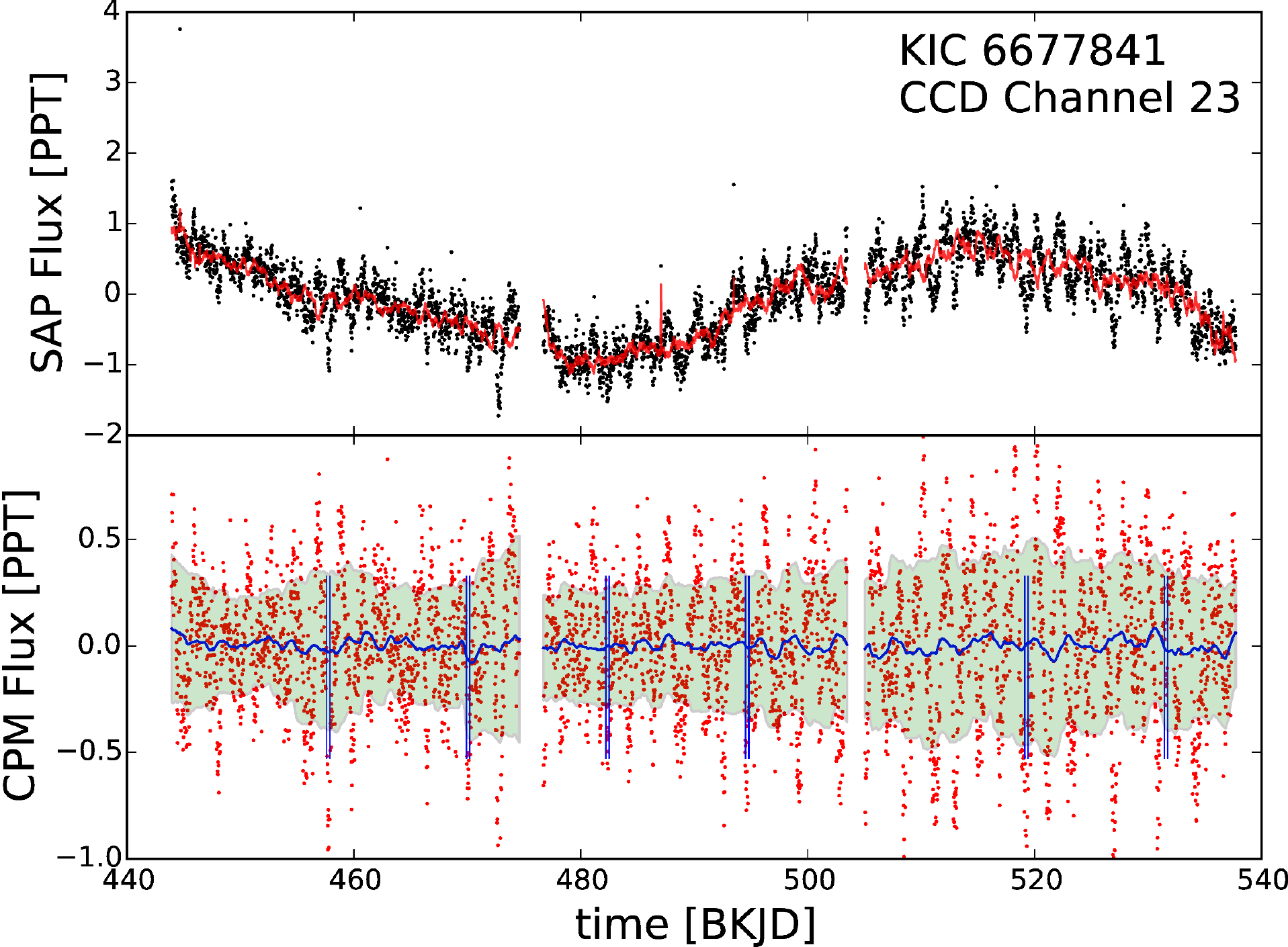}
\end{center}

\caption{
  \label{fluxes} 
  Corrected fluxes using \name, for 9 example stars, spanning the \Kepler\ magnitude (brightness) range. 
  In first column plots, we consider brighs stars with magnitude from 9 to 10.5, second column are stars with magnitude from 11 to 12.5, and the third column are faint stars with magnitude above 13. 
  In the first row, we present some quiet stars. 
  The top panel shows the SAP flux (black) and the \name\ regression (red). 
  The middle panel shows the \name\ flux corrected using the regression, and the bottom shows the PDC flux. 
  One can see that the \name\ flux curve preserves the exoplanet transits (little downward spikes), 
    while removing a substantial part of the variability present in the PDC flux. 
  For the variable stars, we did not present the PDC comparison, 
    since the PDC method preserves intrinsic stellar variability.}
\end{figure}

\begin{figure}[p]
\begin{center}
\includegraphics[width=0.32\textwidth]{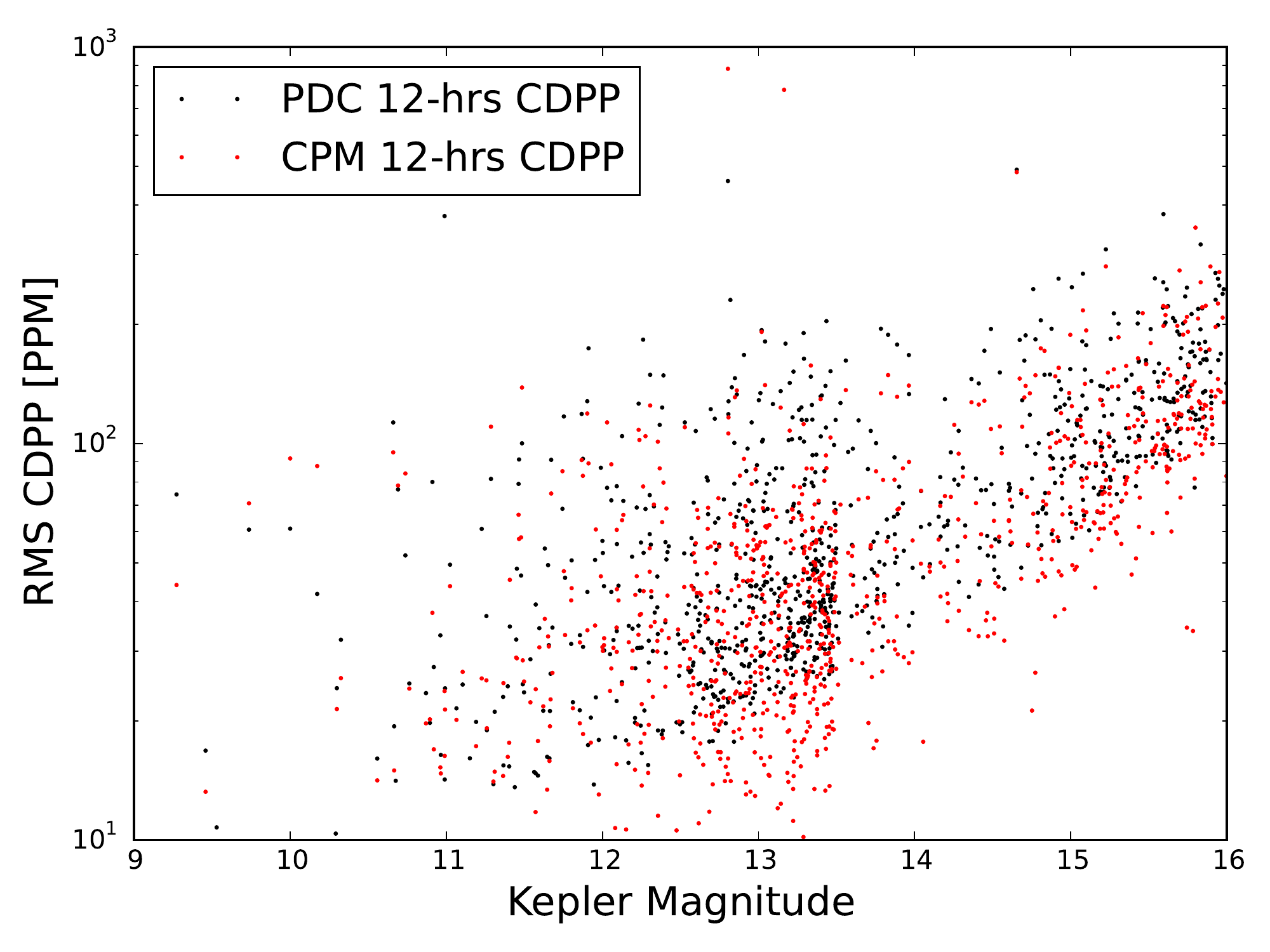}
\includegraphics[width=0.32\textwidth]{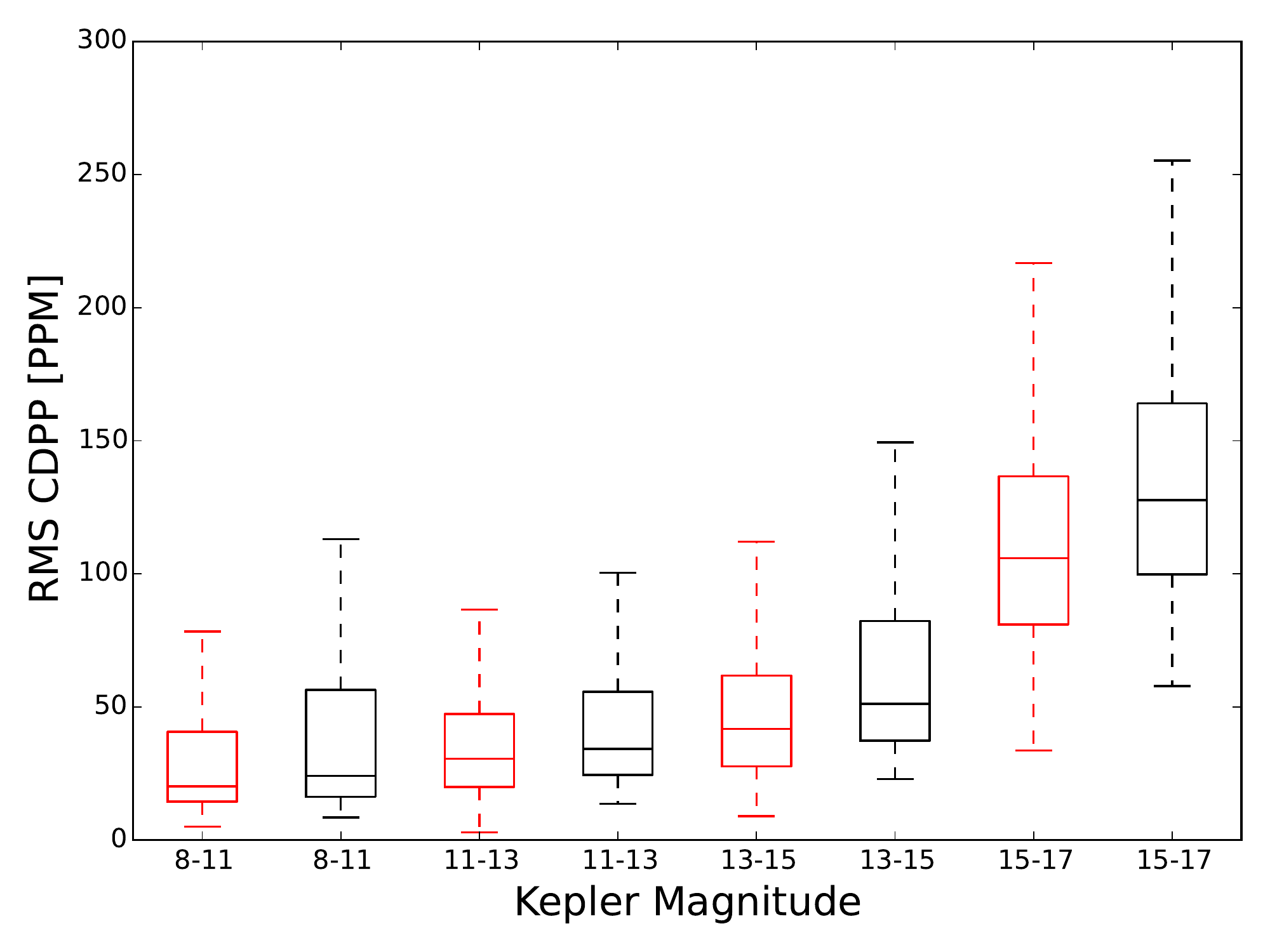}
\includegraphics[width=0.32\textwidth]{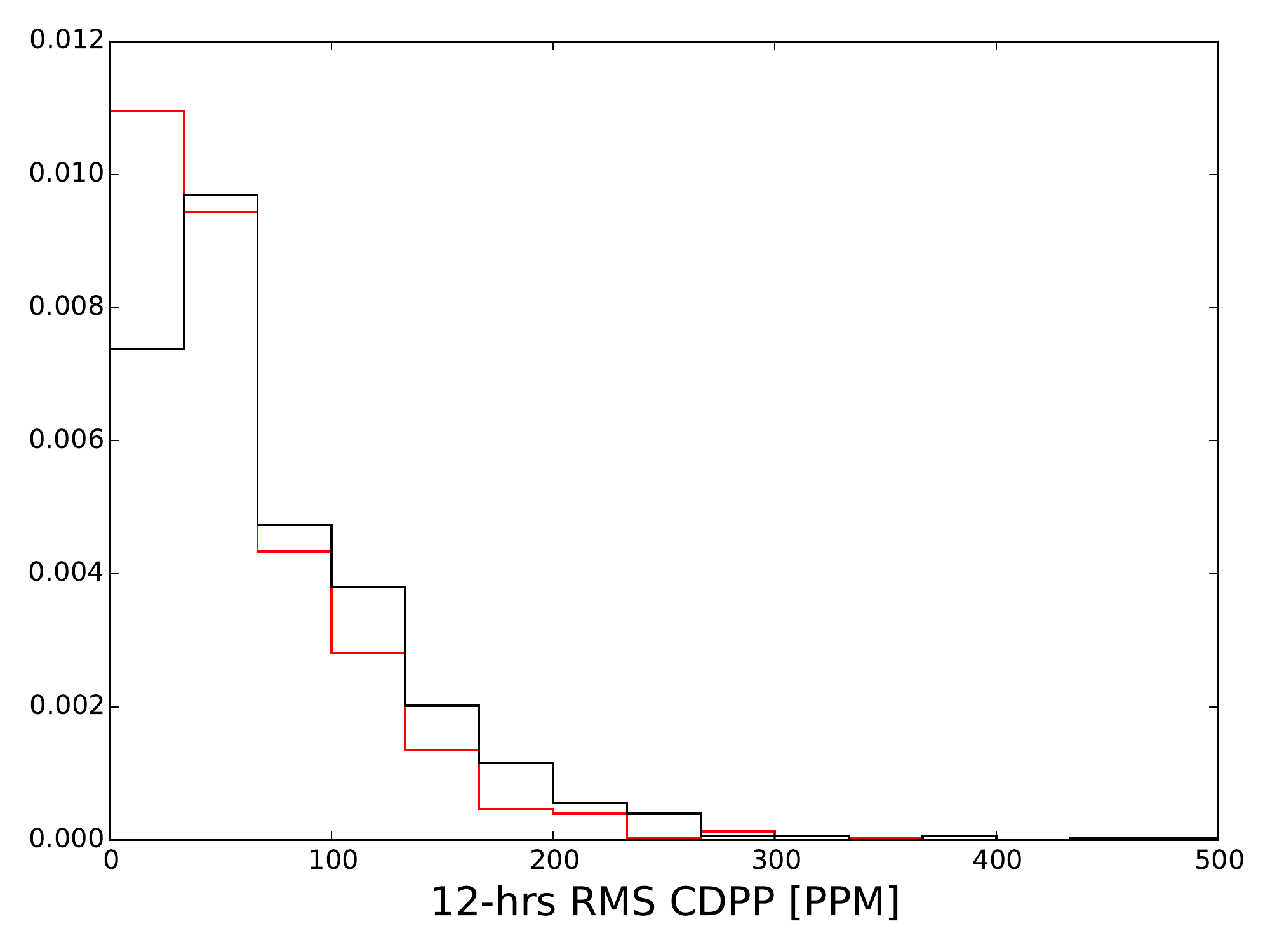}

\includegraphics[width=0.32\textwidth]{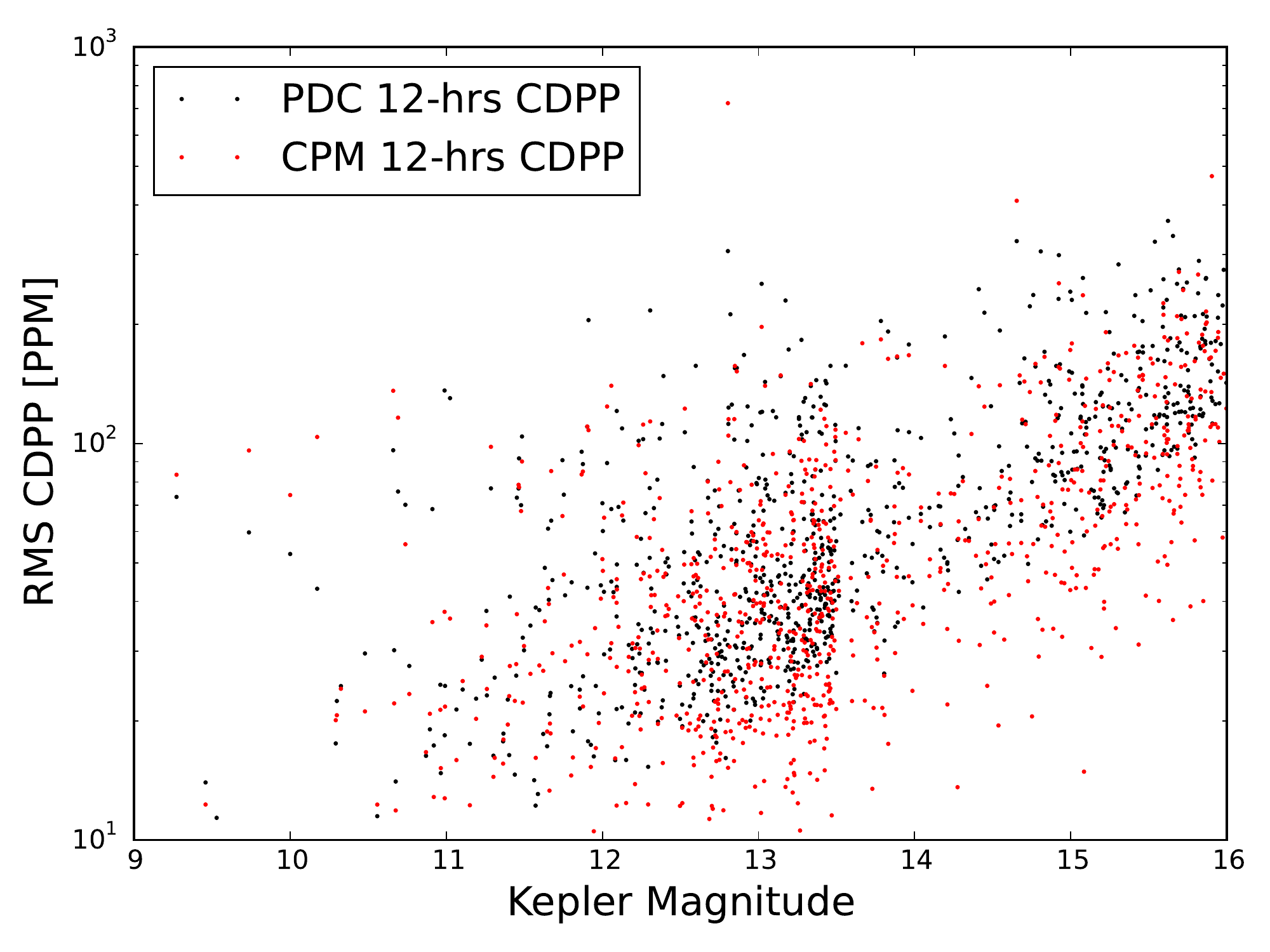}
\includegraphics[width=0.32\textwidth]{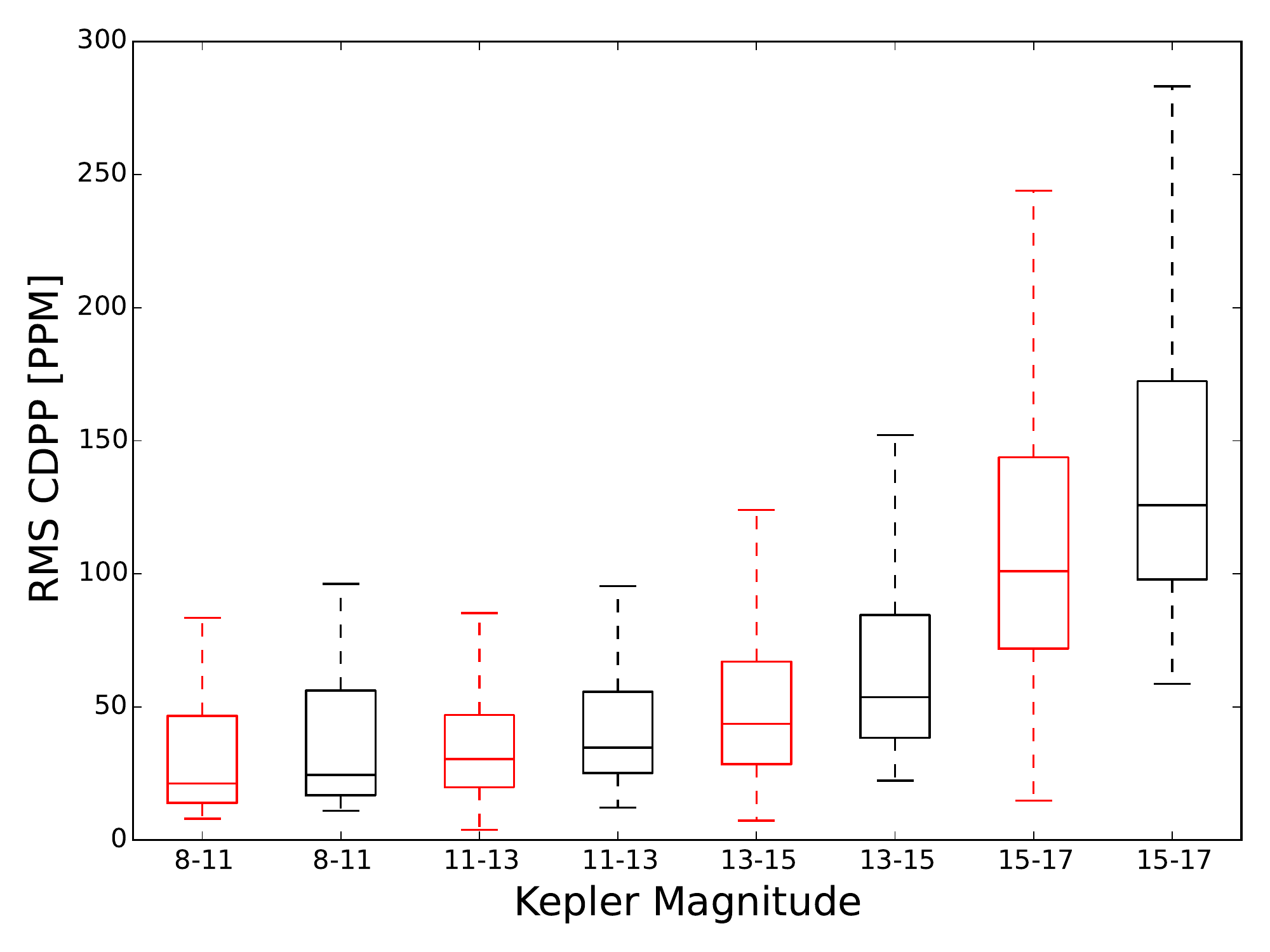}
\includegraphics[width=0.32\textwidth]{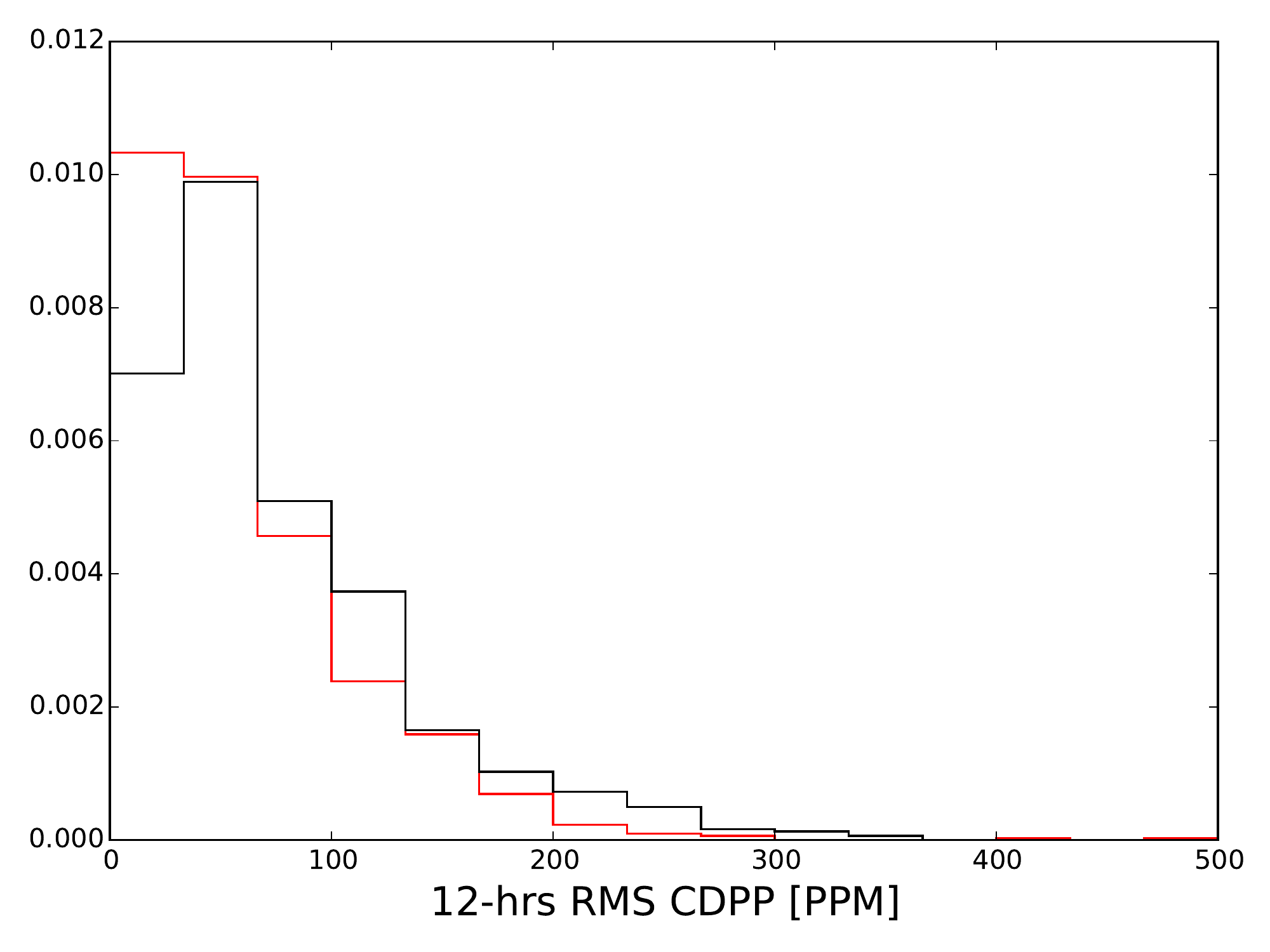}

\includegraphics[width=0.32\textwidth]{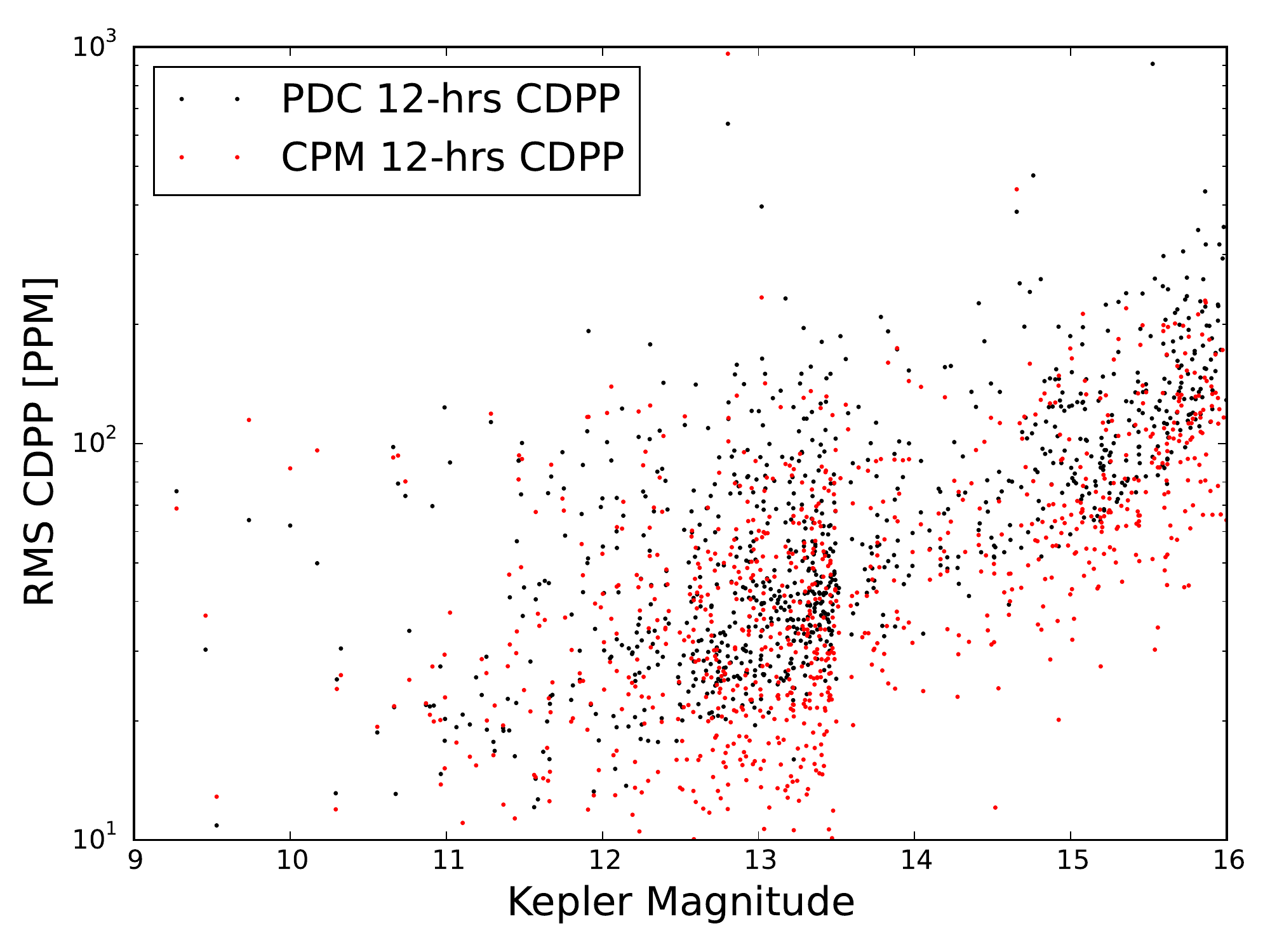}
\includegraphics[width=0.32\textwidth]{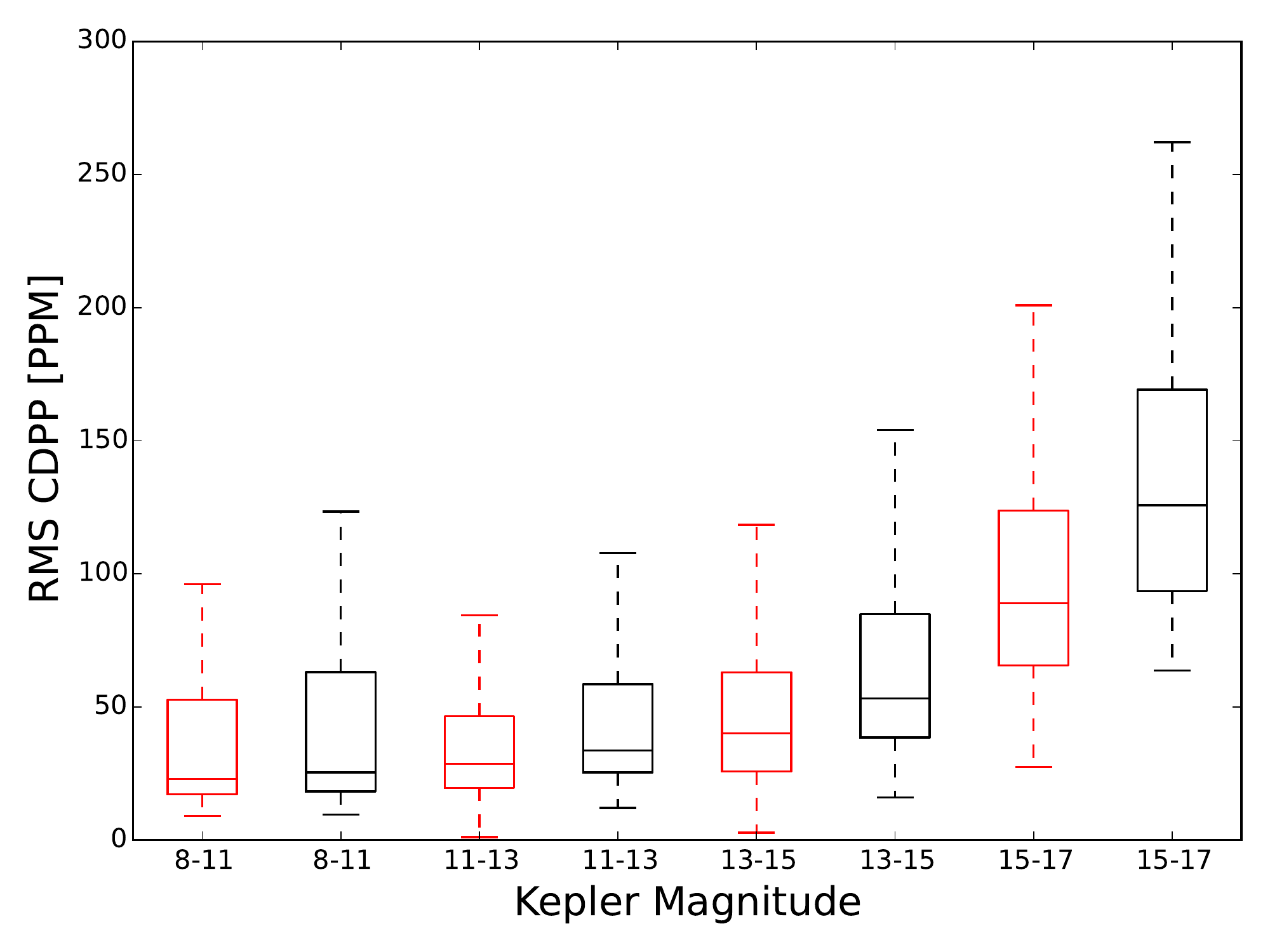}
\includegraphics[width=0.32\textwidth]{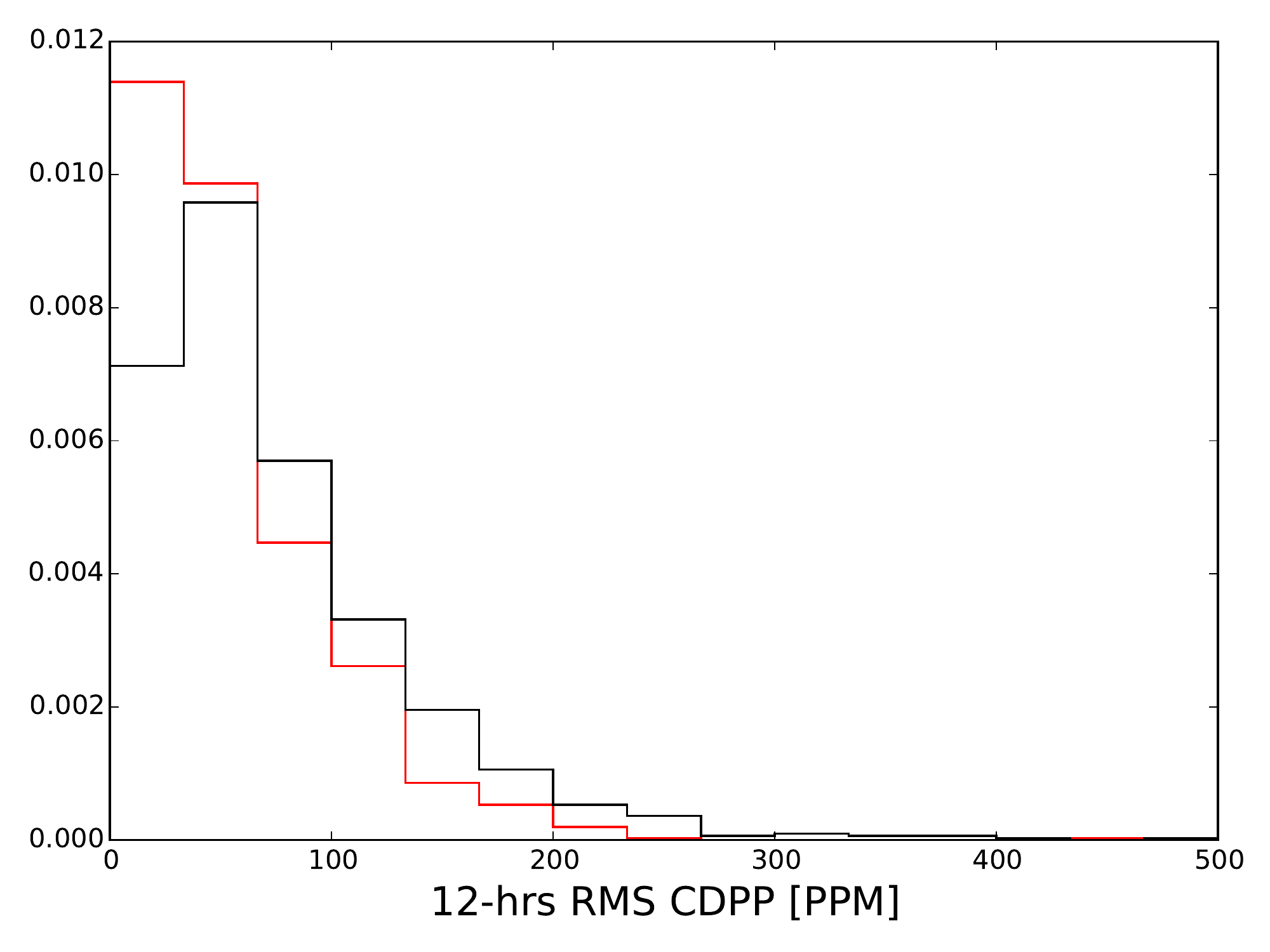}

\includegraphics[width=0.32\textwidth]{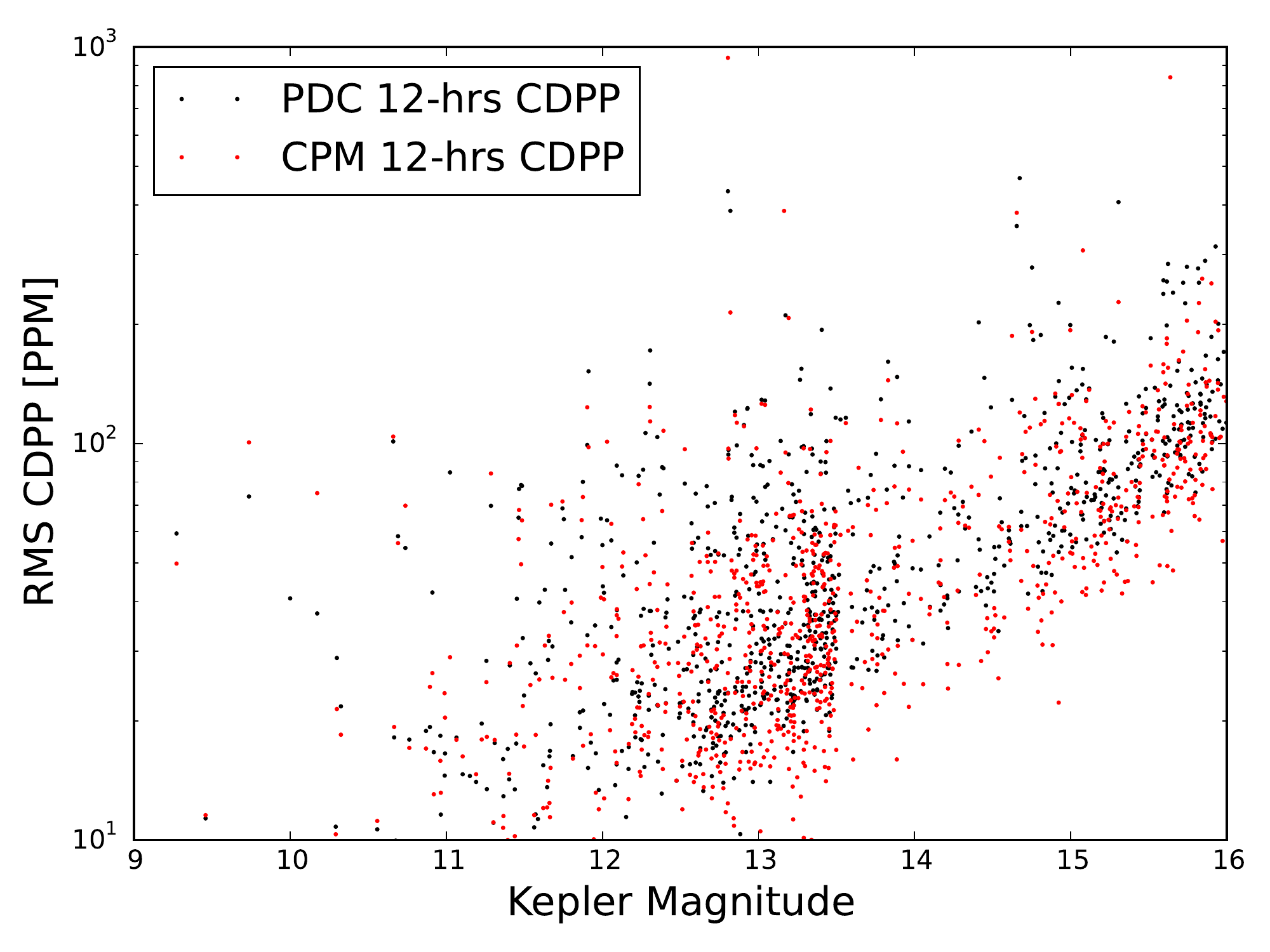}
\includegraphics[width=0.32\textwidth]{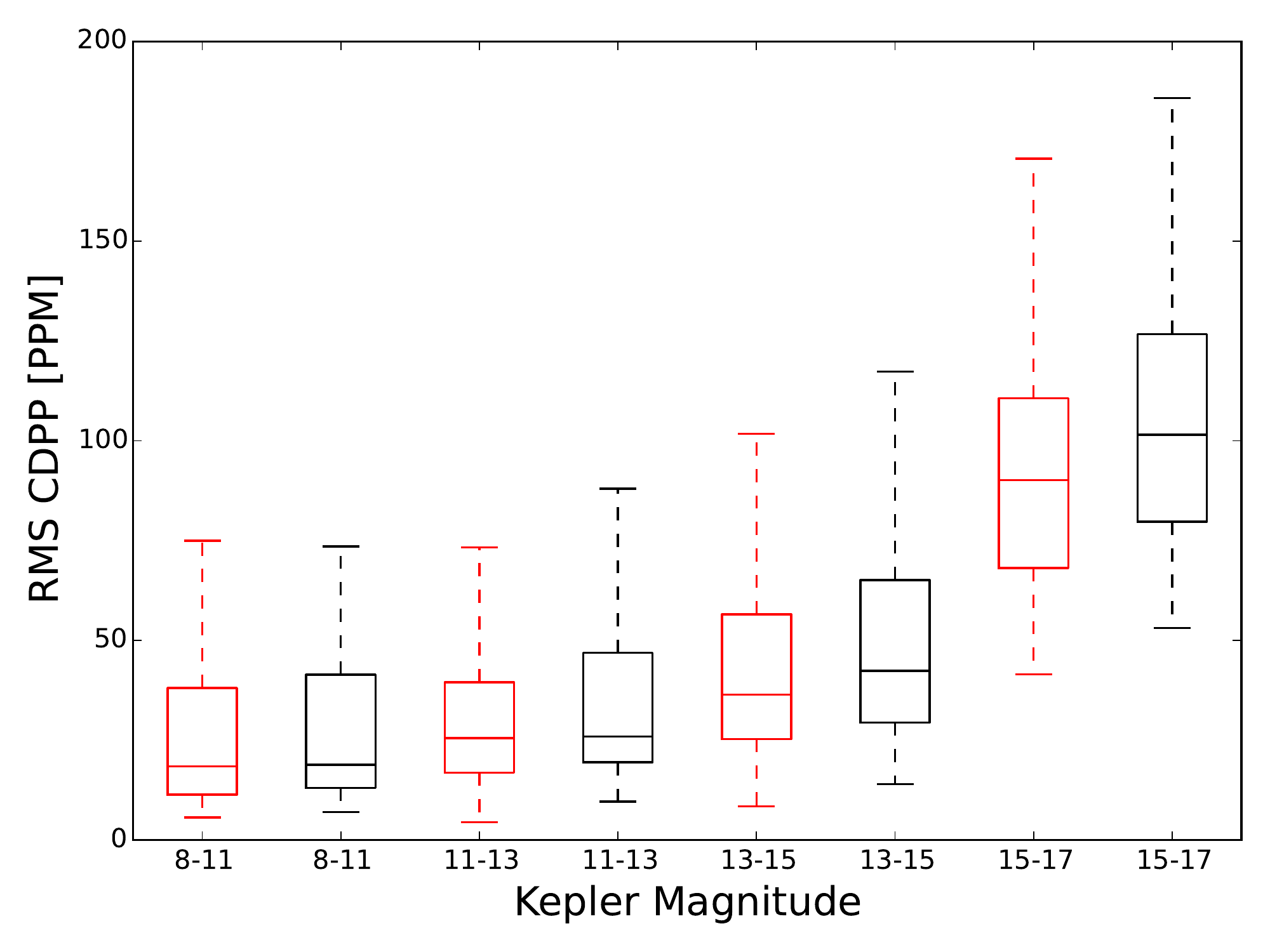}
\includegraphics[width=0.32\textwidth]{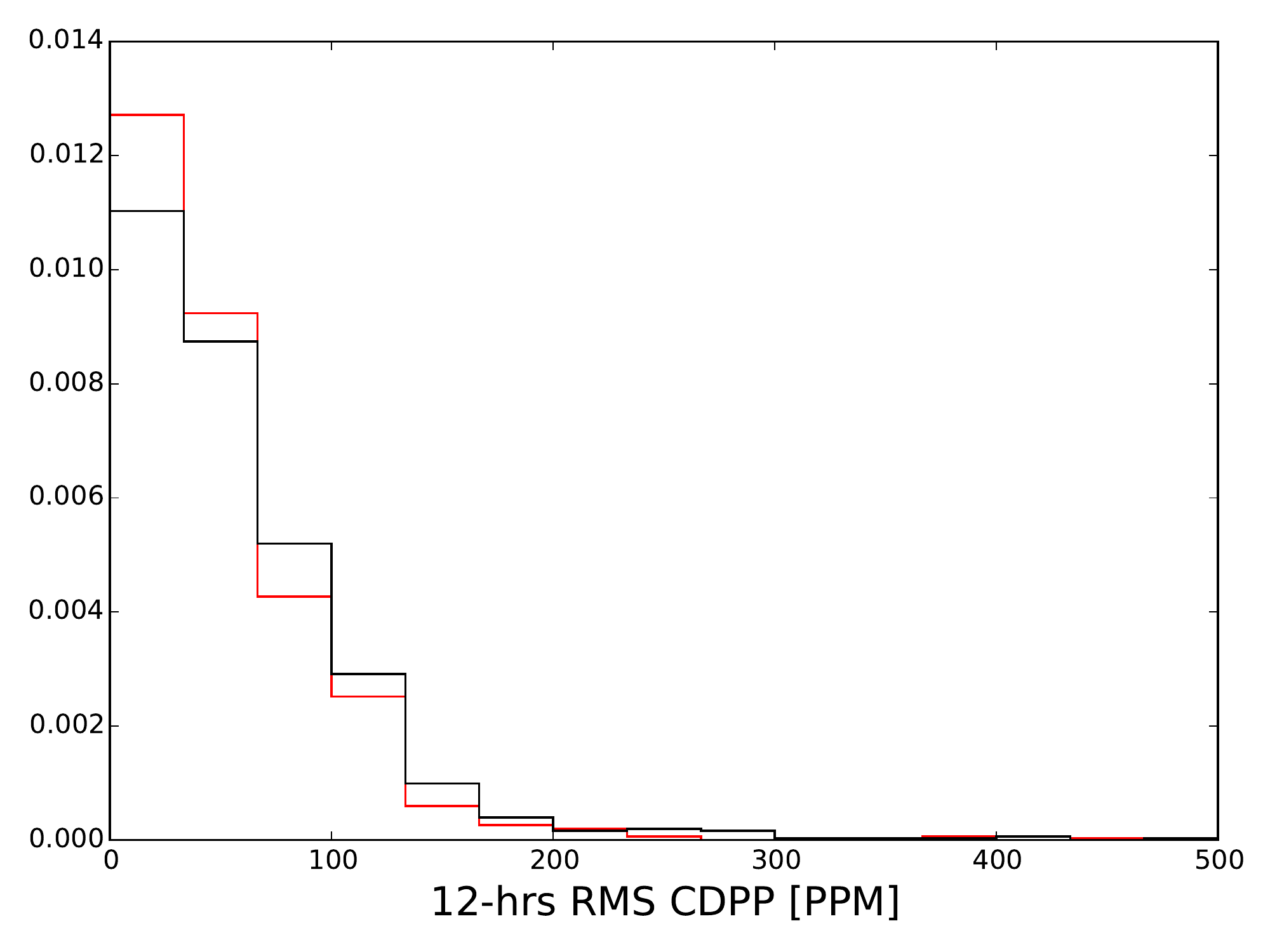}
\end{center}
\caption{
  \label{cdpp} 
  Comparison of the \name\ to the \Kepler\ Presearch Data Conditioning (PDC) method in terms of Combined Differential Photometric Precision (CDPP, calculated with equations 6-8 of section 3.3 in \citealt{cdpp1}).
  Row 1-4 shows the CDPP estimation of 900 stars (500 chosen randomly from the whole \Kepler\ input catalog, and 500 random G-type sun-like stars, the 100 most variable stars are removed) throgh quarter 5 to quarter 8. In each row: 
  \emph{Left} shows our performance (red) vs.\ the PDC performance (black) in a scatter plot, as a function of star magnitude (note that larger magnitude means fainter stars, and smaller values of CDPP indicate higher quality.)
  \emph{Middle} bins the same dataset and shows box plots within each bin, indicating median, top quartile and bottom quartile. 
  The red box corresponds to \name, while the black box refers to PDC. 
  \emph{Right} shows a histogram of CDPP values. 
  Note that the red histogram has more mass towards the left, i.e., smaller values of CDPP, 
    indicating that our method returns the light curves with a lower overall noise level.
}
\end{figure}

The \name\ is applied on several stars with known transit signals but different variability and magnitudes, as shown in Fig.~\ref{fluxes}. 
The \name\ works well on these stars, producing \name\ fluxes 
with low variation and preserving the transit signals.
In addition, \name\ and PDC are compared for on 6 quiet stars (non-variable stars, first two rows of Fig.~\ref{fluxes}). 
The results illustrate that for these quiet stars our approach removes the majority of the variability present in the PDC light curves, while preserving the transit signals. 
To provide a quantitative comparison, we ran \name\ on 1000 stars from the whole Kepler input catalog 
(500 chosen randomly from the whole list, and 500 random G-type sun-like stars), and estimated the Combined Differential Photometric Precision (CDPP,  \citealt{cdpp1} )for the \name\ and PDC.  
CDPP is an estimate of the relative precision in a time window, indicating the noise level seen by a transit signal with a given duration. 
In this paper, the CDPP is calculated by using the equations 6-8 in section 3.3 of \citealt{cdpp1}. The CDPP of both PDC and \name\ is estimated with the same algorithm.
The duration is typically chosen to be 3, 6, or 12 hours. 
We use the 12-hour CDPP metric, since the transit duration of an earth-like planet is roughly 10 hours. 
Before the CDPP is estimated, the variability of the stars is evaluated by median differential variability 
  (MDV, see \citealt{basri2013} for details). 
Based on the variability, the 100 most variable stars in the list are removed, 
  since PDC is not designed to remove stellar variability, 
  and we want to compare \name\ and PDC on the same footing.
Fig.~\ref{cdpp} presents our CDPP comparison of \name\ and PDC, 4 quarters' (quarter 5, 6, 7, 8. Data of quarter 5 are from Kepler data release 21 and other data are from Kepler data release 24---the current version) data are used to show that \name\ has a consistent performance over a whole orbital period of the Kepler photometer (372 days). We also want to emphasize that both \name\ and PDC are presented here, but since PDC is designed to preserve stellar variability, it is not fair to put \name\ and PDC into competition. 
The CDPP comparison is made only to give a relative reference how \name\ can perform according to PDC, the Kepler ``gold standard".

\begin{figure}[p]
\begin{center}
\includegraphics[width=\textwidth]{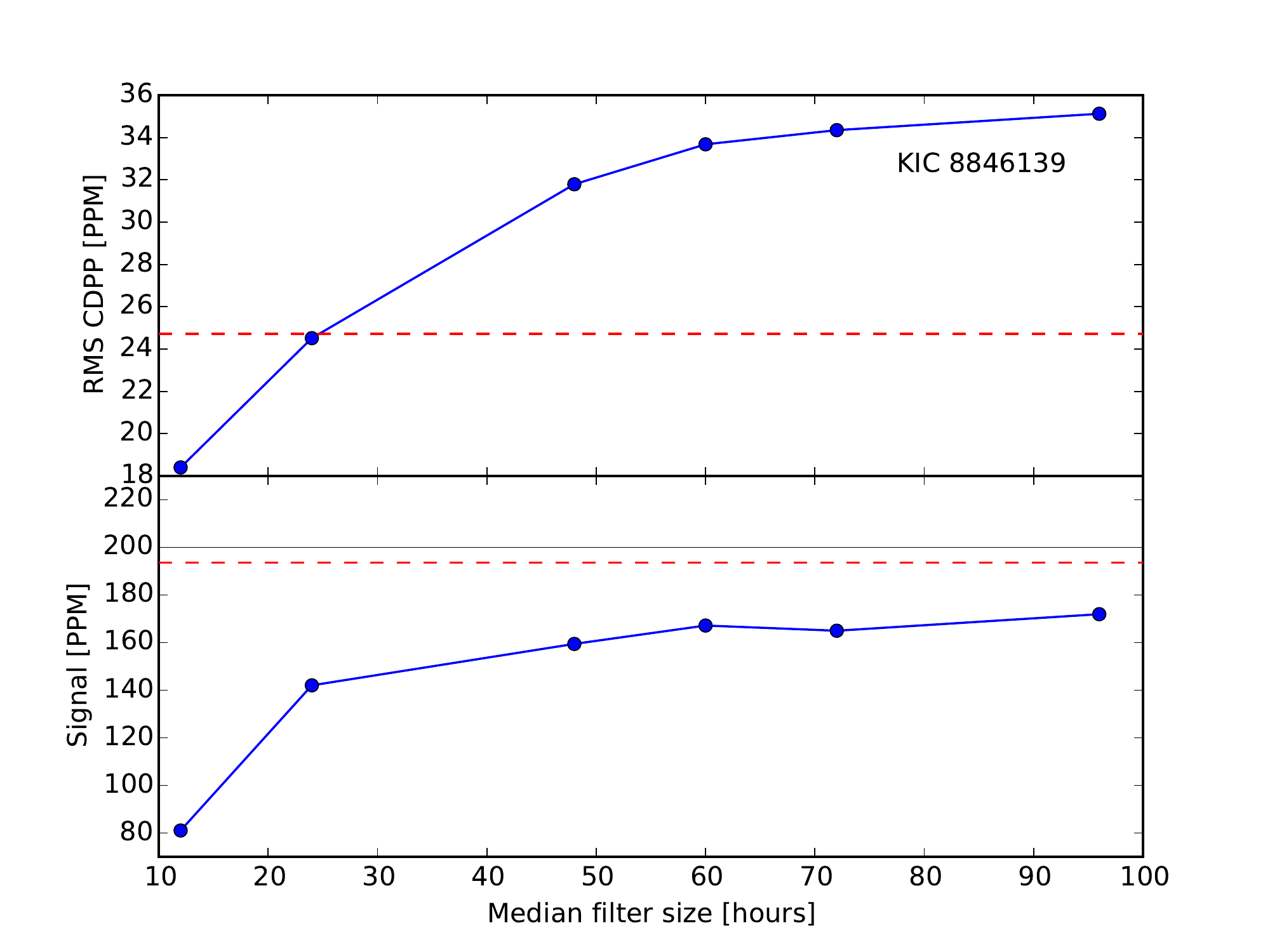}
\end{center}
\caption{
  \label{filter} 
Comparison of the \name\ to the median filter method as described in text with different filter window size. A 200 ppm amplitude box-model signal was injected into the light curve of KIC 8846139 in quarter 5. Both the \name\ and median filter method are applied to the light curve after injection. In the first panel, CDPP estimate of the median filtered light curve is plotted as a function of the filter window size (blue dots) and the red dash line indicates the CDPP level of the \name. In the second panel, the signal strength measured from the median filtered light curve is plotted as a function of the filter window size (blue dots), while the red dash line indicates the signal strength retrieved from the \name\ light curve. The black solid line shows the injected signal strength. Smaller window size reduces the noise level (CDPP) of the median filtered light curve, but distorts the signal more. \name\ outperforms the median filter in terms of signal to noise ratio.
}
\end{figure}

In order to show the overall performance of \name\ for exoplanet search, a comparison between the \name\ and a median filter (a widely used de-trending method) is presented in Fig.~\ref{filter}. This is a more comprehensive comparison, since not only the noise level (CDPP) is evaluated,  but how well both methods preserve signals in the light curve is also under consideration. In the example, a 200 ppm box model signal was injected into a quiet star KIC 8846139 in quarter 5. With the signal-injected light curve,  both methods---median filter and \name---are applied. As shown in Fig.~\ref{filter}, when  the window size is bigger than 24 hours, \name\ performs better both in terms of CDPP and signal strength. As expected, when the window size gets smaller, CDPP decreases,  while the signal strength retrieved from the light curve also falls off. Although the median filter with window size smaller than 24 hours is able to generate cleaner light curves (lower CDPP level) than \name, it largely distorts the signal that we care about.

Imagining an extreme case, light curves generated by a half-hour median filter will have zero CDPP, as well as zero signal strength. If we examined the ratio of the signal strength to the CDPP (signal to noise ratio) in this example, it results that \name\ can always have a signal to noise ratio around 8, while the median filter method can only achieve a ratio around 6 in the best circumstance, when window size is 24 hours. Therefore, the example intuitively exhibits that \name\ is able to generate light curves with low noise level as well as preserve the transit signal.

\section{Discussion}
We have presented a simple yet highly effective method---the \name---that works at pixel-level to remove spacecraft and stellar variability on time-domain imaging. It is based on the causal structure of the \Kepler\ data; It calibrates the \Kepler\ light curves for the purpose of exoplanet search.
In the \name, systematics and stellar variabilities are removed by either fitting with other stars' light curves or auto-regressive components, while transit signals are preserved with a train-and-test framework to control model freedom.
Low variable clean light curves can be produced by \name, which is ideal for planet search. 

Apart from the \name, there exist several other methods that are effective in de-trending light curves.
Methods like the median filter are quite successful in smoothing unwanted features 
  either from spacecraft or intrinsic stellar variability, 
  but they filter everything in the data including the transit signal. 
This is not a big deal for searches for giant planets, since the signal is quite strong. 
However, 
  when it comes to Earth-like planets (only $\backsim 100$\ ppm transit signal around a sun-like star), protecting transit signal from distortion is important.
In comparison, in order to achieve higher precision for Earth-like planet searching, \name\  not only exploits the causal structure of the \Kepler\ data,
  but also effectively (through strong regularization and train-and-test framework) avoids over-fitting the transit signal.
There also exists more sophisticated methods like PDC.

As mentioned earlier in this paper, \name\ is very similar to PDC, in that they both make use of the correlation between lightcurves, which is assumed to be caused by the spacecraft effects (or systematics).
However, the main differences between \name\ and PDC are the following:
One is that \name\ calibrates every single pixel separately, while PDC works at the photometry level.
Pixel-level modelling enables the \name\ to capture more variability, such as 
variations in the centroids and point-spread function from spacecraft pointing, roll, and temperature.
The other reason is that, in the PDC,  only the leading eight principal components of some relative light curves are used.
Although restricting the number of components can prevent over-fitting, it is insufficient to capture all the variabilities in the light curves.
In the \name\, hundreds of stars' light curves (or thousands of pixels) are used to capture relatively complete information of the variability, while strong regularization and train-and-test framework are applied to prevent over-fitting.
Although we present the comparison between \name\ and PDC, 
  we still want to emphasis that PDC is a method intended to preserve stellar variabilities, 
  while the \name\ is optimized for exoplanet searching. 
The comparison is only based on our objective (searches for Earth-like planets), 
  and should not be regarded as standard of a best model for \Kepler\ data.
  
However, apart from the good performance in calibrating the data, the \name\ still has several issues. 
Based on our assumption of the causal structure of the \Kepler\ data, if we turn off the auto-regressive components and keep the predictors from other stars working in the \name, we should be able to still remove the systematics while preserving the intrinsic stellar variability, since there is no reason that these independent stars can be used to predict the stellar variability of the target star.
However, in fact, we can not preserve the stellar variability just by turning off the auto-regressive components. 
The reason is that the light curve is always overfitted to some extent, though in the \name, both strong regularization and train-and-test framework are performed to prevent overfitting. Train-and-test framework is perfect for preventing overfiting within time-scale $\Delta t$ (size of the excluded region) , but it can do nothing for time-scale longer than $\Delta t$. Since there are a lot of long-term trends in the stellar variability, the \name\ must overfit these stellar variabilities with time-scale that is longer than $\Delta t$.
Thus for the time being, we just focus on producing well-calibrated light curves for searching exoplanets, but in long term, we are looking forward to find a method to preserve the stellar variability.

Another issue with the \name\ is that when there are strong transit signals in the light curve, one can find distortion around these signals.
We think this problem is caused by the train-and-test framework.
The train-and-test framework is applied in the \name\ to preserve transit signals.
However, for data points a half-window size away from the transit signals, the prediction of these points in \name\ is influenced by the strong transit signal.
That is,  these points are predicted to be a little bit smaller than they should be, since the transit signal (negative points away from the continuum) makes the model think there is a decreasing trend in this region.
This window edge effect is a trade-off of the train-and-test framework that is crucial for \name, and can not be solved within the scope of the \name.

One possible way to solve this kind of edge effect but still preserve the transit signals is to perform simultaneous fitting for both systematics and transit signal together \citep{dfm}. 
That is,  instead of only modelling the systematics and then subtracting them from the raw data to produce de-trended light curves, a comprehensive model for all components in the data (systematics, stellar variabilities and transits)  can be made. In this kind of model, there is no need for the train-and-test framework, since all transit signals have already been modelled and no de-trended light curve will be produced. 
Planets will be found directly in the model itself. 

Another issue or improvement that can be considered for the \name\ is an improved selection algorithm for predictor pixels. 
In the \name\, a simple selection policy is applied, in which stars closest in magnitude on the same CCD channel are selected. This algorithm might be reasonable, since intuitively the response function of CCD might be similar for stars with similar brightness. 
However, Kepler is a complicated system. Systematics like rolling bands \citep{handbook} may affect only some of the stars. Chances are that ``bad predictors" may be included into the set of predictor stars to distort the \name\ prediction or \name\ might miss some of the ``good predictors", in which case, it can not capture some of the systematics signature. In the current implementation, the \name\ uses a huge set of predictors (4000 pixels from other stars), which ensures that the model will perform well in general, but pixel selection still remains an unsolved issue for the \name.
In this paper, we did not purse optimized selection algorithms, because we want to make the model as simple as possible. But in order to get optimized performance for \name, how to select and rank the predictor stars is an important and unavoidable issue.

As we have mentioned before in this paper,  since the \name\ needs to make predictions for every data point in the light curve, it is quite time-consuming to process the complete Kepler data set with \name. Here we present \name\ runtime estimation based on our experience: For a light curve of duration three months (one quarter), the \name\ takes about 7 hours on a single core (Ivy Bridge x86 64 3.0GHz).

However, despite all these issues, we expect that a data-driven model like our method will enable astronomical discoveries at higher sensitivity on the existing \Kepler\ data as well as on future missions.  
As we know, \Kepler\ \project{K2} \citep{k2} data are suffering from the failure of reaction wheels,  which makes the space craft hard to control and introduces significant systematics. 
Flexible data-driven model \citep{dfm} has already shown its power on the \project{K2} data to find new planets.
Moreover, in 2017, NASA is planning the launch of another space telescope---\project{TESS} (Transiting Exoplanet Survey Satellite, \citealt{tess}), 
  which will perform an all-sky survey for small (earth-like) planets of nearby M stars. 
We think our method can be simply extended for these project to help achieving higher precession and more scientific results.

\acknowledgements
It is a pleasure to thank the whole \Kepler\ Team
  for designing, delivering, and operating a great facility,
  and for making all of the data public, in all its rawest forms, through the MAST interface.
We are also pleased to thank
  Ruth~Angus (Oxford),
  Tom~Barclay (Ames),
  Bekki~Dawson (Berkeley),
  Rob~Fergus (NYU),
  Stefan~Harmeling (Dusseldorf),
  Michael~Hirsch (UCL),
  Dustin~Lang (CMU),
  Benjamin~T.~Montet (Caltech),
  David~Schiminovich (Columbia),
  and
  Jake Vanderplas (UW)
for valuable discussions, input, encouragement, and advice.
This project was partially supported by
  NSF grant IIS-1124794,
  NASA grant NNX12AI50G.
  the Moore Foundation,
  and
  the Sloan Foundation.
This research made use of the NASA Astrophysics Data System.

\clearpage
\bibliography{ms}
\clearpage

\end{document}